\documentclass[a4paper, 11pt]{article}
\pdfoutput=1 % if your are submitting a pdflatex (i.e. if you have
\usepackage{jheppub} % for details on the use of the package, please
\usepackage[T1]{fontenc} % if needed
\usepackage{blkarray}
\usepackage{relsize}
\usepackage{extarrows}
\usepackage{filecontents}

\usepackage[utf8]{inputenc}
\usepackage{amssymb, amsthm, tabto, epigraph, mathtools, dsfont, verbatim, slashed, mathrsfs}
\usepackage{graphicx, wrapfig}
\usepackage{enumitem}
\usepackage{relsize}
\usepackage{subcaption}
\usepackage{csquotes}
\usepackage{hyperref}
\usepackage{physics}
\usepackage{tabularx}
\usepackage[dvipsnames]{xcolor}
\usepackage{color}

\usepackage{amsmath}
\usepackage{empheq}
 
\setlist{nolistsep}
\hypersetup{
    colorlinks=true,
    linkcolor=Mahogany,
    filecolor=magenta,      
    urlcolor=blue,
}

\newlength\dlf  % Define a new measure, dlf

\allowdisplaybreaks

\setlist{nolistsep}
\hypersetup{
    colorlinks=true,
    linkcolor=Maroon,
    filecolor=Maroon,      
    urlcolor=Maroon,
    citecolor=Maroon
}

\newtheorem{theorem}{Theorem}

\theoremstyle{remark}

\theoremstyle{definition}

\usepackage{amssymb}
\usepackage{tikz}
\usetikzlibrary{shapes,arrows,cd,chains,decorations.markings,decorations.pathmorphing,calc}
\tikzset{
->-/.style args={#1rotate#2}{decoration={markings, mark=at position #1 with {\arrow[scale=1.5,rotate = #2 ]{stealth}}}, postaction={decorate}}
}
\tikzset{curve/.style={settings={#1},to path={(\tikztostart)
    .. controls ($(\tikztostart)!\pv{pos}!(\tikztotarget)!\pv{height}!270:(\tikztotarget)$)
    and ($(\tikztostart)!1-\pv{pos}!(\tikztotarget)!\pv{height}!270:(\tikztotarget)$)
    .. (\tikztotarget)\tikztonodes}},
    settings/.code={\tikzset{quiver/.cd,#1}
        \def\pv##1{\pgfkeysvalueof{/tikz/quiver/##1}}},
    quiver/.cd,pos/.initial=0.35,height/.initial=0}
\tikzset{tail reversed/.code={\pgfsetarrowsstart{tikzcd to}}}
\tikzset{2tail/.code={\pgfsetarrowsstart{Implies[reversed]}}}
\tikzset{2tail reversed/.code={\pgfsetarrowsstart{Implies}}}

\newmuskip\pFqskip
\mathchardef\pFcomma=\mathcode`, % keep a copy of the comma

\graphicspath{{./images/}}

\DeclareMathOperator{\Hom}{Hom}

\newcommand{\R}{\mathbb{R}}
\newcommand{\C}{\mathbb{C}}
\newcommand{\Z}{\mathbb{Z}}

\newcommand{\cC}{\mathcal{C}}
\newcommand{\cA}{\mathcal{A}}
\newcommand{\cL}{\mathcal{L}}

\newcommand{\cM}{\mathcal{M}}

\newcommand{\cH}{\mathcal{H}}
\newcommand{\cZ}{\mathcal{Z}}

\renewcommand{\bar}{\overline}

\usepackage{braket}

\usepackage[export]{adjustbox}

\title{\center{Particle-Soliton Degeneracies from \\ Spontaneously Broken Non-Invertible Symmetry}}

\abstract{We study non-invertible topological symmetry operators in massive quantum field theories in (1+1) dimensions. In phases where this symmetry is spontaneously broken we show that the particle spectrum often has degeneracies dictated by the non-invertible symmetry and we deduce a procedure to determine the allowed multiplets. These degeneracies are robust predictions and do not require integrability or other special features of renormalization group flows. We exhibit these conclusions in examples where the spectrum is known, recovering soliton and particle degeneracies.  For instance,  the Tricritical Ising model deformed by the subleading $\mathbb{Z}_{2}$ odd operator flows to a gapped phase with two degenerate vacua. This flow enjoys a Fibonacci fusion category symmetry which implies a threefold degeneracy of its particle states, relating the mass of solitons interpolating between vacua and particles supported in a single vacuum.}

\author[*]{Clay C\'ordova,}
\author[*]{Diego Garc\'ia-Sep\'ulveda,}
\author[*]{and Nicholas Holfester}

\affiliation[*]{Kadanoff Center for Theoretical Physics \& Enrico Fermi Institute, University of Chicago}

\emailAdd{clayc@uchicago.edu}
\emailAdd{dgarciasepulveda@uchicago.edu}
\emailAdd{nholfester@uchicago.edu}

\begin{document}

\maketitle
%\tableofcontents
%\newpage

\section{Introduction} \label{Introduction}

One of the classical applications of symmetry principles in quantum field theory (QFT) is to organize the spectrum of particle excitations.  Indeed, it is  Poincar\'{e} symmetry that defines the concept of particle to begin with as a certain type of representation in Hilbert spaces.  Internal global symmetries further refine this picture by grouping particles into multiplets of equal masses transforming as representations of the symmetry group.  The main goal of this work is to generalize these results to the case of more general, higher symmetry principles \cite{Gaiotto:2014kfa}.  These algebraic symmetries are dictated by fusion categories \cite{Frohlich:2009gb,Carqueville:2012dk, Bhardwaj:2017xup, Chang:2018iay} and we will show that they enforce novel degeneracies in the particle spectrum of QFTs.

The primary setting of our analysis is QFTs in (1+1) dimensions where the higher symmetries of interest are realized as topological line operators. There has been much recent work analyzing generalized global symmetries in these models with nearly all attention paid to fixed points of the renormalization group (RG) \cite{Thorngren:2019iar, Lin:2019hks, Thorngren:2021yso, Lin:2022dhv,  Chang:2022hud, Burbano:2021loy,  Choi:2023vgk, Diatlyk:2023fwf,Bashmakov:2023kwo,Chen:2023jht, Lin:2023uvm}.  For instance, in (1+1)-dimensional rational conformal field theory (CFT) there are topological Verlinde lines \cite{Verlinde:1988sn, Petkova:2000ip, Drukker:2010jp, Gaiotto:2014lma} in one-to-one correspondence with the primaries of the chiral algebra. These lines form a fusion category (see e.g.\ \cite{etingof_tensor_2015}), and dictate much of the structure of the conformal field theory.  Denoting the distinct lines as $a,b,c, \cdots$, a characteristic feature is a fusion algebra:
\begin{equation}
	a \otimes b = \sum_c N_{ab}^{c} \, c, \,  \quad N_{ab}^c \in \Z_{\geq 0}.
\end{equation}
In particular, the appearance of multiple terms in the sum on the right-hand side implies that the operators $a,b,c,$ etc.\ typically do not have inverses.  For this reason they are often referred to as non-invertible to emphasize the contrast with ordinary group-like symmetries, which are always invertible. Such non-invertible symmetries have a myriad of applications: they can constrain the spectrum of scaling dimensions \cite{Lin:2022dhv, Lin:2023uvm}, encode features such as dualities of chiral algebras and CFTs \cite{Cordova:2023jip, Choi:2023vgk, Diatlyk:2023fwf, Lin:2019hks, Thorngren:2021yso}, and also have a rich interplay with the study of boundary conditions in CFT \cite{Choi:2023xjw, Collier:2021ngi}.  There are also emerging examples of irrational CFTs with non-invertible symmetry \cite{Cordova:2023qei}, and a corresponding interplay with quantum gravity when these CFTs are interpreted as the worldsheet of a dynamical string \cite{Kaidi:2024wio, Heckman:2024obe}.  

The other class of RG fixed points that have been widely studied are topological quantum field theories (TQFTs). These systems are characterized by a finite set of topological local operators and their operator product expansion. They enjoy a (multi)fusion category symmetry and in a precise sense that we discuss below, this generalized global symmetry completely dictates the correlators of such theories \cite{Inamura:2021wuo, Huang_2021, Moore:2006dw}.  CFTs, TQFTs and their fusion category symmetries are natural starting points to understand more general QFTs \cite{Komargodski:2020mxz, Copetti:2024rqj, Bhardwaj:2023fca, Perez-Lona:2023djo, Robbins:2021ibx}. Indeed a typical picture of an RG flow is a CFT at short distances perturbed by a relevant operator.  Often the long distance dynamics is gapped in which case in the extreme IR the theory is governed by a TQFT encoding the distinct vacua:
\begin{equation}
    \text{CFT} \xlongrightarrow{\text{RG}} \text{TQFT}.
\end{equation}
Assuming that the relevant operator generating the flow preserves some of the non-invertible symmetry of the CFT, then the entire RG flow has a fusion category symmetry, $\mathcal{C},$ that can be used to constrain the QFT.  This is the setting of our analysis.  

An initial question about the structure of such $\mathcal{C}$ symmetric flows is what can be said a priori about the vacua.  For instance we can ask if the IR can support a unique $\mathcal{C}$ symmetric vacuum or if instead some of the symmetry must be spontaneously broken.  This is the question of anomalies of the fusion category symmetry.  As first discussed in \cite{Chang:2018iay, Chang:2022hud}, when the quantum dimensions of the simple lines are not integers then some of the fusion category symmetry must be spontaneously broken leading to degenerate vacua, i.e.\ a non-trivial IR TQFT.  More generally, the associators of the fusion category can also be anomalous again implying some spontaneous symmetry breaking
\cite{Zhang:2023wlu} (higher-dimensional analogs of anomalies of non-invertible symmetry have also been explored in \cite{Choi:2021kmx, Choi:2022zal, Apte:2022xtu,Kaidi:2023maf, Antinucci:2023ezl, Cordova:2023bja}, and analogs on the lattice are deduced in \cite{OBrien:2017wmx,Aasen:2020jwb,Eck:2023gic,Cheng:2022sgb, Seiberg:2023cdc,Seiberg:2024gek}.)  This point of view of utilizing $\mathcal{C}$ to understand the general structure of a massive phase can be viewed as a generalized Landau paradigm for fusion category symmetries \cite{Bhardwaj:2023fca, Bhardwaj:2023idu, Bhardwaj:2024qrf, Bhardwaj:2023bbf, Zhao:2020vdn, Wen_2019}.  

 It is also natural to ask how fusion category symmetry constrains RG flows beyond just the structure of the vacua.  For instance, in \cite{Komargodski:2020mxz} it was found that non-invertible symmetry provides novel selection rules in effective field theory, while in \cite{Copetti:2024rqj} fusion categories were shown to imply unusual versions of crossing symmetry in the S-matrix. Moreover, in the context of the lattice, \cite{Aasen:2020jwb} constructed models where non-invertible symmetry is explicitly shown to enforce degeneracies between excited states which, in a solvable limit, resemble particles and kink-antikink pairs. In this vein, our primary question addressed in this work is to ask if fusion category symmetry can generally constrain the massive particle spectrum of a QFT.  We will see that the answer to this question is positive and our main result is a procedure derived in Section \ref{sec:genstructure} below, which elucidates a general method to determine the implied degeneracies.  

\begin{figure}[t]
        \centering
        \includegraphics[scale=0.35]{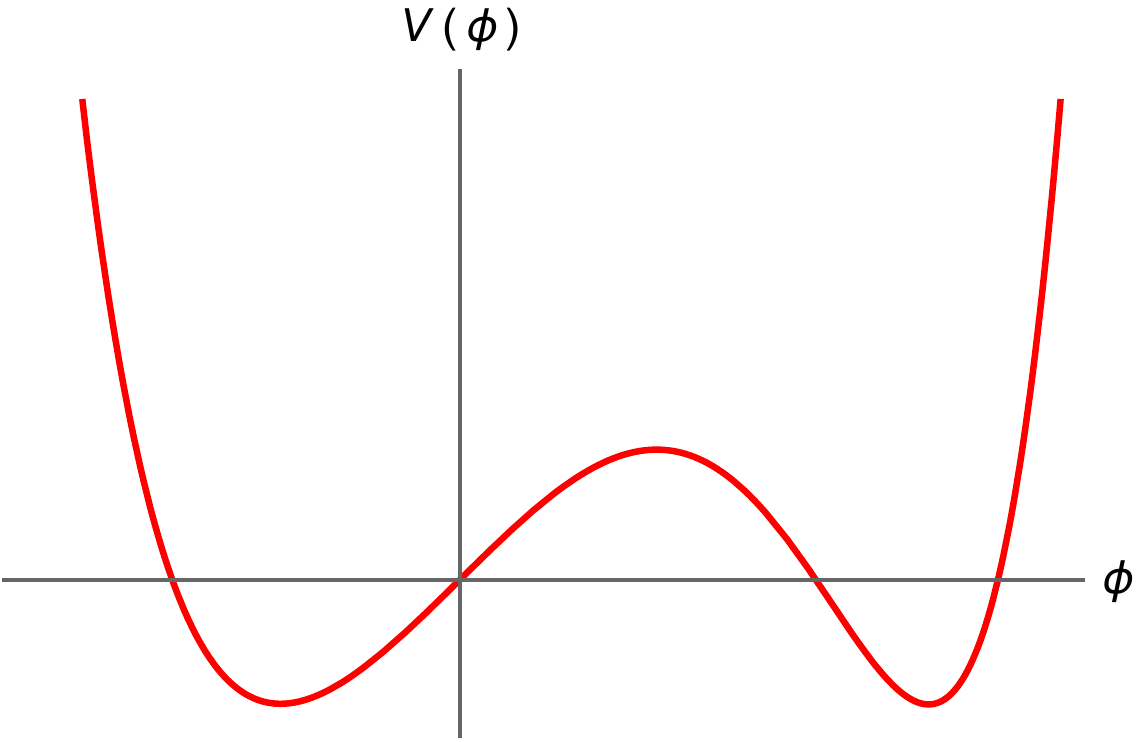} 
        \caption{Schematic depiction of the associated Landau-Ginzburg potential of the Tricritical Ising model deformed by the $\phi_{2,1}$ primary operator. Note in particular the absence of a $\mathbb{Z}_{2}$ symmetry in the potential.} \label{Tricriticalphi21plot}
\end{figure}

As an invitation to our results let us consider as an example the Tricritical Ising model deformed by the relevant operator $\phi_{2,1}$ of scaling dimension $\Delta=7/8$.  This flow preserves a single non-invertible topological line $W$ which obeys a Fibonacci fusion rule:
\begin{equation} \label{Z2TYIntro}
    W \otimes W = 1 + W .
\end{equation}
The massive particle spectrum of this flow was first obtained numerically in \cite{Lassig:1990xy} and by bootstrap considerations in \cite{Zamolodchikov:1990xc, Colomo_1992_1, Colomo_1992_2}.  To gain intuition about this system, we can realize it in terms of a Landau-Ginzburg flow.\footnote{Strictly speaking, the minimal model flow and the Landau-Ginzburg model need not preserve the same topological line operators along the flow. Therefore, we showcase the potential to introduce the spectrum found in \cite{Lassig:1990xy} and never use it directly in our calculations. \label{footnoteLGmodel}}  The ultraviolet is then a scalar field theory with dynamical variable $\phi$ and potential (see Figure \ref{Tricriticalphi21plot}):
\begin{equation}
    V(\phi) = \phi^{6}-10 \lambda^{3}\phi^{3}+12 \lambda^{5}\phi.
\end{equation}
Here $\lambda\geq 0$ is a coupling constant.  For vanishing $\lambda$ the theory flows to the Tricritical Ising CFT. Meanwhile for non-zero $\lambda$ the theory has two degenerate minima at:
\begin{equation}\label{minimaintro}
\phi_{W}=-\lambda \zeta^{-1}, \hspace{.3in}\varphi_{1}=\lambda \zeta,
\end{equation}
where $\zeta$ is the golden ratio, and describes the same long-distance physics as the deformed minimal model.  We observe that in contrast to more standard double well potentials, 
the system does not exhibit a $\mathbb{Z}_{2}$ symmetry exchanging the vacua.  Instead, as suggested by the labelling of the minima in \eqref{minimaintro}, these vacua should be viewed as exchanged by the spontaneously broken non-invertible symmetry $W$.

The Hilbert space of states splits into sectors, $\mathcal{H}_{ij}$, where the vacuum at $-\infty$ is $i$ and the vacuum at $+\infty$ is $j$.  In the model at hand this leads to the sectors:
\begin{equation}
    \mathcal{H}=\mathcal{H}_{1 , 1}\oplus \mathcal{H}_{W,W} \oplus \mathcal{H}_{1, W} \oplus \mathcal{H}_{W, 1}.
\end{equation}
The sectors $\mathcal{H}_{1, W}$ and $\mathcal{H}_{W, 1}$ each support a stable single particle state, which semi-classically one may view as a soliton.  These are a particle-antiparticle pair and hence have the same mass $m_{s}$.  According to \cite{Lassig:1990xy,Zamolodchikov:1990xc,Colomo_1992_1}, apart from the solitons, there is also a single particle state in the sector $\mathcal{H}_{W,W}$ of mass $m_{p}$. In the context of integrability, these particles that do not interpolate between different vacua are known as ``breathers.''  Since these states do not interpolate between different vacua their stability is more mysterious.  Moreover, strikingly, \cite{Lassig:1990xy, Zamolodchikov:1990xc,Colomo_1992_1} found that these states are exactly degenerate in mass:
\begin{equation}\label{massintro}
    m_{s}=m_{p},
\end{equation}
despite the fact that there is no ordinary internal symmetry present to enforce this result.  Instead as we will derive below, the degeneracy \eqref{massintro} is enforced by the Fibonacci symmetry \eqref{Z2TYIntro}.  Thus the triplet of states, \{soliton, anti-soliton, particle\}, can be viewed as the minimum multiplet of the non-invertible symmetry.  In particular, this prediction is robust and does not require integrability or other special features of an RG flow.  Instead it holds for any model with spontaneously broken Fibonacci symmetry. 

To derive the kind of conclusions described above, we proceed in Section \ref{sec:genstructure} to systematically unpack the necessary algebraic structure of fusion category symmetry in (1+1)d QFTs.  While much of this structure is mathematically known, a key conceptual hurdle that we must address is how to utilize the $\mathcal{C}$ symmetry when space is $\mathbb{R}$ as opposed to a more standard compact spatial slice such as $S^{1}$.  This is a necessary complication, since the concept of a particle utilizes the Poincar\'{e} symmetry present only in infinite space.  Our approach to this problem is to view $\mathbb{R}$ as a long distance limit of a segment with specified boundary conditions.  Physically, these boundary conditions are the infrared ground states of the theory determined by the IR TQFT along the flow. Mathematically they are encoded by a so-called module category of $\mathcal{C}$, as explored for instance in \cite{Moore:2006dw,Bhardwaj:2017xup,Huang_2021,Choi:2023xjw}. We elucidate the algebraic data of these boundaries and junctions allowing us to determine how the fusion category $\mathcal{C}$ acts on different sectors of the Hilbert space.  This may be viewed as an open sector analog of the higher-representation theory explored in \cite{Bartsch:2022mpm, Bartsch:2022ytj,Lin:2022dhv, Bhardwaj:2023wzd,Bhardwaj:2023ayw, Bartsch:2023wvv, Copetti:2023mcq}. We also explain how this boundary data can be determined via Lagrangian algebras of a (2+1)d symmetry TQFT \cite{Fuchs:2002cm, Fuchs:2003id, Fuchs:2004dz, Fuchs:2004xi, Gaiotto:2014kfa, Gaiotto:2020iye, Apruzzi:2021nmk,Freed:2022qnc, Chatterjee:2022kxb, Inamura:2023ldn, Kaidi:2022cpf, Bhardwaj:2023bbf, Brennan:2024fgj, Antinucci:2024zjp, Bonetti:2024cjk, Apruzzi:2024htg, Apruzzi:2023uma, DelZotto:2024tae} that encodes the fusion category symmetry and its action of the physical theory by inflow.

In Section \ref{sec:MinModelFlows}, we apply this machinery to models where the particle spectrum is known via integrability \cite{Zamolodchikov:1990xc, Zamolodchikov:1991vh}.  In addition to the flow described above, we also consider unitary minimal models $M_{n}$ deformed by the $\phi_{1,3}$ operator.  There we show that the non-invertible symmetry implies a multiplet of $2(n-2)$ degenerate single particle states consistent with the predictions from integrability. As above, this degeneracy is robust and uses only the fusion category symmetry, and it is not necessary to solve for the exact dynamics.

%%%%%%%%
%%%%%%%%

\section{Fusion Categories, Module Categories, and Degeneracies} \label{sec:genstructure}

In this section we present a working approach to analyzing particle degeneracies enforced by non-invertible symmetries in $(1+1)d$ QFTs with boundary conditions. In Section \ref{subsec:background} we review the underlying algebraic structure of fusion categories and their module categories which physically encode the bulk topological line defects and their topological junctions on boundaries. In Section \ref{subsec:open_sect} we introduce the class of operators defined by stretching a bulk line between boundaries and phrase the study of the degeneracies they enforce in terms of kernels and cokernels. We then present a sufficient condition for such kernels and cokernels to vanish and hence for the existence of degeneracies enforced by a broad class of such maps. In Section \ref{subsec:massive_qft} we specialize this analysis to massive QFTs on $\R^2$ to study stable particle degeneracies enforced by the non-invertible symmetry of the theory. We argue that the module category of clustering boundary conditions is determined by the infrared TQFT of the theory and review how it may be studied using Lagrangian algebras of the corresponding $(2+1)d$ TQFT.

\subsection{Relevant Background}\label{subsec:background}

\subsubsection{Fusion Categories (Bulk)} 

To warm up and fix notation we briefly review the necessary structure of fusion categories and their module categories. For more details see \cite{Bhardwaj:2017xup, Chang:2018iay, Kitaev_2006, Komargodski:2020mxz, Huang_2021, Kitaev_2012, etingof_tensor_2015}. Throughout, all equalities are understood as equalities that hold inside correlation functions.\footnote{More carefully, when inserted on a spacetime manifold, topological junctions come with an ordering \cite{Chang:2018iay}. Diagrammatically we will suppress this from our notation with the implicit choice of ordering being clear in context. When specifying junction labels this ordering is made explicit.}

In this paper we will consider QFTs in $(1+1)d$. In this dimension, $0$-form symmetries are topological line defects. Such line defects can both support and be joined by topological local operators. Denoting two topological lines by $a,b$, the vector space of topological local operators joining them is denoted $\Hom(a,b)$. A line is called \textit{simple} if the only topological local operators it supports are multiples of the identity operator. Under operator product expansion (OPE) topological operators joined by the same line can be fused.

Algebraically the collection of line defects and their topological local operators assemble into a \textit{fusion category} \cite{Carqueville:2012dk,Frohlich:2009gb,Bhardwaj:2017xup,Chang:2018iay}.\footnote{Implicit in this statement is the assumption that the theory has a unique topological local operator. This is the case for QFTs constructed from perturbing a CFT with unique vacuum, which we will consider later. The generalization to multiple topological local operators is mathematically captured by multi-fusion categories.} The data of the fusion category $\cC$ consists of a set of simple, oriented topological lines $\{a,b,\dots\}$ which can be fused 
\begin{equation}
	a \otimes b = \sum_c N_{ab}^{c} \, c, \,  \quad N_{ab}^c \in \Z_{\geq 0},
\end{equation}
vector spaces of topological junctions $j$,
\begin{equation}
    \includegraphics[width=4.7cm, valign=m]{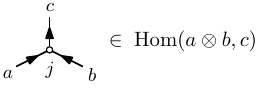},
\end{equation}
and an $F$-symbol defined by the diagram
\begin{equation}\label{eq:fusion_6j}
    \includegraphics[width=8cm, valign=m]{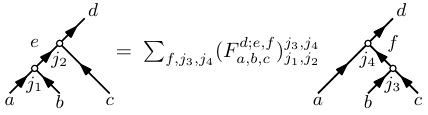}
\end{equation}
with respect to a choice of basis on junction vector spaces. A fusion category $\mathcal{C}$ has the additional property that there is a unique identity line $\mathbf{1}$ which is simple. Physically, the simplicity of $\mathbf{1}$ reflects the theory having a single topological local operator. This is the correct setting for our target application where $\cC$ is the fusion category of topological lines preserved along an RG flow from a CFT with unique vacuum. 

The $F$-symbol is an invertible linear map and is required to satisfy the pentagon equation. This ensures that all decompositions of $n$-point junctions into $3$-point junctions are equal \cite{Chang:2018iay}. Schematically, the pentagon equation is given by the commutative diagram
\begin{equation}\label{eq:fusion_pentagon}
    \includegraphics[width=6.5 cm, valign=m]{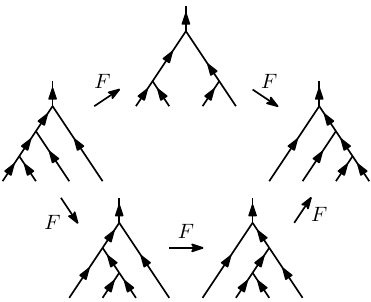}.
\end{equation}
Concretely, for simple multiplicities of fusions (the case we will need) this is a system of polynomial equations
\begin{equation}
    \sum_{h}F_{a,b,c}^{g;f,h}F_{a,h,d}^{e;g,i}F_{b,c,d}^{i;h,j} = F_{f,c,d}^{e;g,j}F_{a,b,j}^{e,f,i}.
\end{equation}
A property of central importance for later application is that solutions to this equation (after modding out those related by changes of basis) do not admit continuous deformations, a result known as ``Ocneanu rigidity'' \cite{ocneanu_rig,etingof_tensor_2015}.

In $\mathcal{C}$, all topological lines are finite sums (with positive integer coefficients) of simple lines. Relatedly, the dimensions of junction spaces of simple lines $\Hom(a\otimes b,c)$ are given in terms of fusion coefficients
\begin{equation}
    \textrm{dim}\Hom(a\otimes b, c) = N_{ab}^c.
\end{equation}
An additional useful fact is that distinct simple lines cannot be joined by a topological local operator\footnote{Distinct meaning ``non-isomorphic''. That is, there does not exist an invertible topological local operator joining $a$ and $b$.}
\begin{equation}
    \Hom(a,b) = 0, \quad a \ncong b.
\end{equation}

The orientation reversal of a line defines its dual
\begin{equation}
	\includegraphics[width=2cm, valign=m]{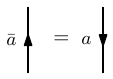}.
\end{equation}
Up to the presence of an orientation-reversal anomaly, a self dual line $a = \bar{a}$ can be represented as a line without an arrow.\footnote{The orientation-reversal anomaly is algebraically captured by the Frobenius-Schur indicator which we will in general allow to be non-trivial. This has the benefit of allowing for the removal of any non-trivial isotopy anomaly. For further discussion of this relation see \cite{Chang:2018iay, Huang_2021}.}
The fusion of a line and its dual contains the identity line and moreover there exists a canonical junction vector in $\Hom(a\otimes \bar{a}, \mathbf{1})$ corresponding to ``forgetting'' the identity line:
\begin{equation}\label{eq:id_junction}
    \includegraphics[width=4cm, valign=m]{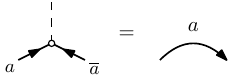}. 
\end{equation}
For all expressions involving $a,\bar{a},$ and $\mathbf{1}$ we always use this junction. For simple lines this property characterizes duals, that is
\begin{equation}\label{eq:dual_id}
	a \otimes b = \mathbf{1} + \dots \quad \Leftrightarrow \quad b = \bar{a}.
\end{equation}
In the special case that $a \otimes \bar{a} = \mathbf{1}$ the line is called \textit{invertible}. We will also need that there is a natural pairing of junction vectors of simple lines
\begin{equation}\label{eq:fusion_pairing}
    \includegraphics[width=4 cm, valign=m]{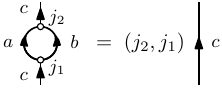}
\end{equation}
with $(j_2,j_1) \in \C$. Geometrically this corresponds to shrinking the loop to a local operator on $c$ which is then a multiple of the identity operator on $c$. For the case of $c = \mathbf{1}$ and the junctions \eqref{eq:id_junction}, this defines the quantum dimension of $a$:\footnote{Viewing these diagrams as drawn in local patches of the spacetime manifold, this definition implicitly chooses the isotopy anomaly (if present) to be cancelled \cite{Chang:2018iay}.}
\begin{equation}\label{eq:dimension}
    \includegraphics[width=2.3cm, valign=m]{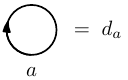}.
\end{equation}
This captures the action of the topological line on the single topological local operator of the QFT.

To simplify the discussion, in the following we will assume that the fusions of lines have simple multiplicities 
\begin{equation}
	a \otimes b = \sum_c N_{ab}^c c, \quad N_{ab}^c \in \{0,1\}.
\end{equation}
This fixes non-zero junction vector spaces to be one dimensional and so only one vector is needed to be specified for each junction. This simplifies the indexing of the $F$-symbol
\begin{equation}
	(F_{a,b,c}^{d;e,f})_{j_1,j_2}^{j_3,j_4} \mapsto F_{a,b,c}^{d;e,f} \in \C.
\end{equation}
The case of general multiplicities can be recovered straight forwardly. For later use we note that there is a basis of junctions such that the local fusion is given in terms of dimensions \cite{Kitaev_2006}
\begin{equation}\label{eq:fusion_dimensions}
    \includegraphics[width=5cm, valign=m]{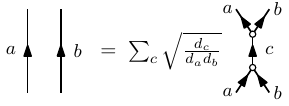}.
\end{equation}
Finally, we recall the centrally important property that the fusion category $\cC$ is a renormalization group (RG) flow invariant since its data does not admit continuous deformations.

\subsubsection{Module Categories (Boundary)}
When placing a QFT on manifolds with boundary there can be a non-trivial interplay between bulk topological lines and boundary conditions. Topological lines can both act on boundary conditions by fusing with them and end transversely on the boundary, joining two boundary conditions. This defines a localized operator at the junction on the boundary. While in general such an operator need not be topological, we will restrict attention to the topological ones with two primary motivations. Heuristically, topological data provides candidates for RG invariants, which can then be used to understand a theory along a flow.\footnote{This intuition is formalized by the theorem that module categories of fusion categories do not admit deformations, another instance of Ocneanu rigidity \cite{etingof_tensor_2015}.} The second motivation is our immediate interest in studying the organization of particle spectra in a massive QFT. Since topological data commutes with continuous spacetime transformations, it provides a natural candidate for enforcing particle degeneracies. Indeed, the essential goal of this paper is to initiate a study of this idea. 

To help orient the reader we outline our trajectory. Below we review that this topological data algebraically assembles into a module category, meaning that every $(1+1)d$ QFT defines a module category from its boundary data. In section \ref{subsec:open_sect} we study what can be learned about states in a QFT given knowledge of its module category. In Section \ref{subsec:massive_qft} we consider the related, but distinct, problem of how one can deduce the module category of a massive QFT. 

To proceed we first specify our notation. Following \cite{Huang_2021}, we label boundary conditions with lines oriented so that the empty theory (which we also color gray) is to the right.\footnote{More carefully, this convention will define the notion of a \textit{left} module category. A \textit{right} module category is defined by the opposite orientation convention. The distinction, however, is not of immediate consequence since for each left module category there is a naturally associated right module category and vice versa. Physically, this reflects that locally, each is related by a rotation.} We will also follow the convention that letters early in the roman alphabet label bulk topological lines, while letters in the center of the alphabet label boundary conditions. Throughout the following $\cC$ will denote the fusion category describing the bulk topological defect lines. 

As mentioned above, boundary conditions of a QFT may be interfaced by topological local operators. For two boundary conditions $m,n$, the vector space of such operators is denoted $\Hom(m,n)$.\footnote{In standard abuses of notation, we will use the same notation for structures in $\mathcal{C}$ and the module category $\cC$. The distinction should always be clear from the context. We will also do this for the multiplication operation $\otimes$ in the following.} A boundary condition is called \textit{simple} if the only topological local operators it supports are multiples of the identity. A closely related notion is decomposability. Like bulk topological lines, boundary conditions can be added in correlation functions. A boundary condition $m$ is \textit{indecomposable} if it is not expressible as a sum of other boundary conditions with non-negative integer coefficients.

Algebraically, topological junctions on boundaries form a \textit{module category} \cite{Kitaev_2012,Bhardwaj:2017xup, Choi:2023xjw,ostrik2001module, etingof_tensor_2015, Huang_2021}. The data of a $\cC$-module category $\cM$ consists of a set of indecomposable boundary conditions $\{m,n,\dots\}$ with which bulk lines can fuse
\begin{equation}
	a \otimes m = \sum_n \tilde{N}_{am}^n n, \quad \tilde{N}_{am}^n \in \Z_{\geq 0},
\end{equation}
vector spaces of topological boundary junctions $j$,
\begin{equation}
    \includegraphics[width=5cm, valign=m]{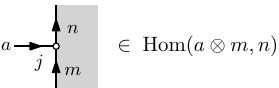},
\end{equation}
and an ``$\tilde{F}$-symbol'' defined by the following diagram
\begin{equation} \label{eq:mod_6j}
    \includegraphics[width=9cm, valign=m]{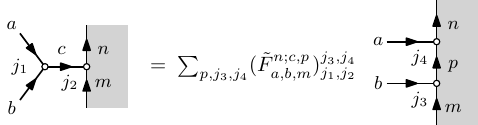}
\end{equation}
for a particular basis. The sum \eqref{eq:mod_6j} is over indecomposable boundary conditions. We emphasize that the boundaries themselves are not required to be topological. The following discussion applies to topological junctions, which non-topological boundary conditions may support. Moreover, one does not need to consider the collection of all boundary conditions but instead may study a subcategory.

The module category $\tilde{F}$-symbol $\tilde{F}$ satisfies a pentagon coherence equation analogous to that of a fusion category, again ensuring different decompositions of $n$-point junctions into $3$-point junctions are equal. Schematically it is given by the commutative diagram
\begin{equation}
    \includegraphics[width=6.2cm, valign=m]{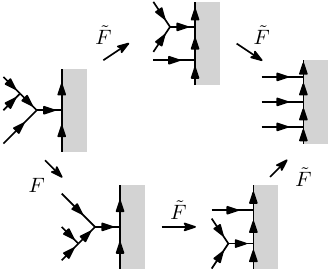}.
\end{equation}
Concretely, for the case of simple multiplicities (our target application) this is the system of polynomial equations
\begin{equation}
    \sum_{f} F_{a,b,c}^{d;e,f} \tilde{F}_{a,f,m}^{n;d,r} \tilde{F}^{r;f,s}_{b,c,m} = \tilde{F}_{e,c,m}^{n;d,s}\tilde{F}_{a,b,s}^{n;e,r}.
\end{equation}
As in the case of the $F$-symbol for a fusion category, the $\tilde{F}$-symbol (modding out changes of basis transformations) is rigid: it does not admit deformations \cite{etingof_tensor_2015}.

For junctions containing the bulk identity line, there is a canonical vector corresponding to ``forgetting'' it
\begin{equation}
    \includegraphics[width=3.4cm, valign=m]{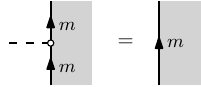}.
\end{equation}
For diagrams involving $m,m$ and $\mathbf{1}$ we will always use this junction. To simplify discussion, in the following we take junction vector spaces $\Hom(a\otimes m, n)$ with $a$ simple and $m,n$ indecomposable to be one dimensional. In particular this requires that all indecomposable boundaries are simple, which will be the case in our target application. The more general case follows straightforwardly.

A straightforward example of a $\cC$-module category $\cM$ that will be of later use is the \textit{regular $\cC$-module category} $\cM_{reg}$. $\cM_{reg}$ contains the same data as $\cC$: simple boundary conditions are labelled by simple bulk lines, the fusion of bulk lines with boundary conditions is given by the bulk fusion
\begin{equation}
	\tilde{N}_{ab}^c = N_{ab}^c \, ,
\end{equation}
boundary junctions are given by bulk line junctions, and the $\tilde{F}$-symbol is that of $\cC$:
\begin{equation}
	\tilde{F}_{a,b,c}^{d;e,f} = F_{a,b,c}^{d;e,f}.
\end{equation}
We emphasize, however, that these are distinct algebraic structures: the boundary conditions in $\cM_{reg}$ cannot be fused. Physically this is clear: $\cM_{reg}$ specifies interfaces of a theory with the empty theory, while $\cC$ specifies interfaces of a theory with itself. Later, we will see such module categories appear in deformations of unitary diagonal minimal models.

\subsection{Open Sectors}\label{subsec:open_sect}
In $(1+1)d$, QFTs are quantized on $1$-manifolds, of which there are topologically four:
\begin{equation}
	S^1, \qquad [0,1], \qquad (-\infty,1],  \quad \mathrm{and} \quad \R.
\end{equation}
On $\cH(S^1)$ bulk topological lines and topological junctions act by ``lassoing'' (see \cite{Chang:2018iay}) and furnish a representation of the so-called tube algebra of the bulk fusion category $\cC$ over all twisted sectors \cite{Lin:2022dhv}. Quantization on the remaining manifolds requires an understanding of boundary conditions. This is clear for the compact manifold $[0,1]$. For the non-compact manifolds $(-\infty,1]$ and $\R$ boundary conditions are needed at infinity to select finite energy states. For example, this is the case in massive QFTs where states are required to resemble clustering vacua at $\pm \infty$. We now consider what can be learned about states in a QFT from knowledge of its module category of boundary conditions. In what follows, we draw diagrams on finite intervals, where the location of the boundaries may be interpreted as spatial $\pm \infty$ to recover the non-compact case. All bulk lines, bulk junctions, and boundary junctions are taken to be topological unless otherwise stated. Our later examples of interest are unitary QFTs and so throughout we take $\cC$ to be unitary.\footnote{A unitary fusion category comes with an additional reflection map that defines an inner product on junctions. The $F$ and $R$ symbols (in some basis) are required to be unitary with respect to these inner products (see Appendix \ref{app:lag_alg} for the definition of the $R$ symbol). For further discussion see \cite{Kitaev_2006, Huang_2021}.}

Let $\cH_{m,n}$ denote the Hilbert space quantized on a spatial $1$-manifold $M_1$ having two boundary conditions $m$ and $n$\footnote{We recall again that in general such boundary conditions are allowed to be non-topological.}
\begin{equation}
	\includegraphics[width=4.2cm, valign=m]{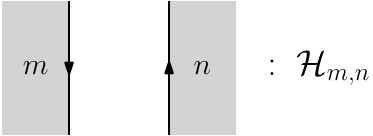}.
\end{equation}
Here (and in the following diagrams) time is chosen to run upwards. In open sectors, a bulk line can be extended between boundaries with topological junctions defining a map between sectors with different boundary conditions
\begin{equation}\label{eq:map_def}
    \includegraphics[width=7.7cm, valign=m]{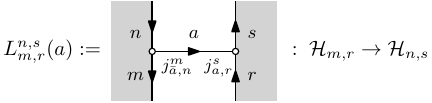},
\end{equation}
where we have temporarily restored junction labels $j_{\overline{a},n}^m,j_{a,r}^s$ to emphasize that such choices are part of the defining data of the map. We would like to understand how such maps organize the states between the $\cH_{n,m}$ sectors.

\subsubsection{Degeneracies}
Because the lines and their boundary junctions under consideration are topological, the associated maps are equivariant over spacetime symmetries in each sector. In particular,
\begin{equation}\label{eq:equivar}
	H_{r,s} \, L_{m,n}^{r,s}(a) =  L_{m,n}^{r,s}(a) \, H_{m,n}
\end{equation}
with $H_{m,n}$ the Hamiltonian on the $\cH_{m,n}$ Hilbert space. Note that equivariance replaces the usual commutator condition since these maps relate different Hamiltonians in different Hilbert spaces.

Take $\ket{\psi} \in \cH_{m,n}$ an eigenstate of energy $E$ and suppose there is a non-zero operator $L_{m,n}^{r,s}(a)$. From \eqref{eq:equivar}
\begin{equation}
    H_{r,s} L_{m,n}^{r,s}(a)\ket{\psi} =  L_{m,n}^{r,s}(a) H_{m,n}\ket{\psi} = E (L_{m,n}^{r,s}(a) \ket{\psi}).
\end{equation}
Note that it does not immediately follow that there is a state of energy $E$ in $\cH_{r,s}$: one must check that $L_{m,n}^{r,s}(a) \ket{\psi} \neq 0$. Similarly, in the converse direction an eigenstate $\ket{\phi} \in \cH_{r,s}$ of energy $E$ only implies a corresponding state of energy $E$ in $\cH_{m,n}$ if there exists a state $\ket{\varphi}$ such that $\ket{\phi} = L_{m,n}^{r,s}(a)\ket{\varphi}$. More succinctly, both conditions are respectively:
\begin{equation}
    \ket{\psi} \notin \ker L_{m,n}^{r,s}(a), \quad \ket{\phi} \notin \textrm{coker} L_{m,n}^{r,s}(a).
\end{equation}
The possible non-triviality of kernels and cokernels is intimately related to the non-invertibility of the bulk topological lines.

The total Hilbert space of the QFT will decompose over different indecomposable boundary condition sectors:
\begin{equation}
	\cH = \bigoplus_{m,n}\cH_{m,n}.
\end{equation}
The above considerations show that topological lines attached to topological boundary junctions may potentially enforce energy degeneracies between sectors of the total Hilbert space $\cH$. To understand if there are such degeneracies, however, the kernels and cokernels of such maps must be understood. The simplest possible case to study is when kernels or cokernels can be shown to vanish, in which case degeneracies are immediate. We proceed now with demonstrating a sufficient condition for this. To do so we will first need an explicit understanding of the compositions of this class of maps.

\subsubsection{Compositions}
Here we will take both bulk lines and boundaries to be simple, with the general case following by linearity. The composition of maps of the type \eqref{eq:map_def} is determined by the data of the bulk category of lines and the boundary module category. It is computed from the following manipulations:
\begin{equation}\label{eq:composition}
    \includegraphics[width=11.5cm, valign=m]{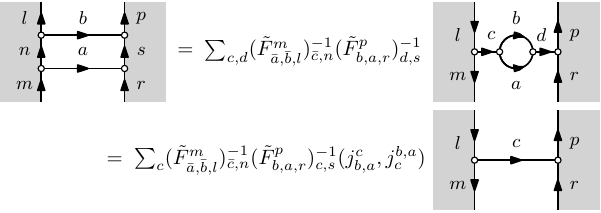}
\end{equation}
Unpacking this expression, $(\tilde{F}_{b,a,r}^p)^{-1}$ denotes the matrix inverse of the module category $F$-symbol \eqref{eq:mod_6j} with $(\tilde{F}_{b,a,r}^p)^{-1}_{c,s}$ indexing the specific matrix element in the $c,s$ channel. Here $(j_{b,a}^c,j_c^{b,a})$ is the pairing \eqref{eq:fusion_pairing} of the (implicit) internal junctions $j_{b,a}^c$ and $j_c^{b,a}$. Less diagrammatically, this says
\begin{equation}\label{eq:composition}
	L_{n,s}^{l,p}(b) \circ L_{m,r}^{n,s}(a) = \sum_c  C_{m,n,l}^{r,s,p}(a,b;c) L_{m,r}^{l,p}(c),
\end{equation}	
with the coefficient given above:
\begin{equation}\label{eq:comp_coeff}
	C_{m,n,l}^{r,s,p}(a,b;c) = \left(\tilde{F}^{m}_{\bar{a},\bar{b},l}\right)^{-1}_{\bar{c},n}\left(\tilde{F}^{p}_{b,a,r}\right)^{-1}_{c,s} (j_{b,a}^c,j_{c}^{b,a}).
\end{equation}

\subsubsection{Kernels and Cokernels}
The most direct way of showing that kernels or cokernels vanish is by constructing a left or right inverse respectively. Using \eqref{eq:composition} we check explicitly if we can construct left or right inverses from a bulk line and boundary junctions.

Focusing on the case of kernels, the kernel of $L_{m,r}^{n,s}(a)$ is required to vanish if there exists a map $L_{n,s}^{m,r}(b)$ satisfying
\begin{equation}
	L_{n,s}^{m,r}(b) \circ L_{m,r}^{n,s}(a) = \alpha \, L_{m,r}^{m,r}(\mathbf{1}) \equiv \alpha \, \mathrm{Id}_{\mathcal{H}_{m,r}}, \quad \alpha \in \C\setminus\{0\} \, ,
\end{equation}
where $\mathrm{Id}_{\mathcal{H}_{m,r}}$ is the identity morphism in the Hilbert space $\mathcal{H}_{m,r}$. The proportionality condition is allowed since junctions may be rescaled. Note that this requires the sum in the composition truncate at the identity channel. Restricting focus to simple lines and simple boundary conditions, the above condition can only hold if $b = \overline{a}$, as a non-zero identity channel in the partial fusions requires
\begin{equation}
    \includegraphics[width=4.6cm, valign=m]{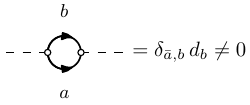} \, ,
\end{equation}
and such a non-zero junction exists if and only if $b = \overline{a}$. Therefore, in this case the condition reduces to that there are boundary junctions satisfying
\begin{equation}\label{eq:ker_condition}
	C_{m,n,m}^{r,s,r}(a,\bar{a};c) = \delta_{1,c} C_{m,n,m}^{r,s,r}(a,\bar{a};\mathbf{1}), \quad  C_{m,n,m}^{r,s,r}(a,\bar{a};\mathbf{1}) \neq 0,
\end{equation}
and so reduces the problem to computing
\begin{equation}\label{eq:vanish_ker_condition}
	C_{m,n,m}^{r,s,r}(a,\bar{a};c) = \left(\tilde{F}^{m}_{\bar{a},a,m}\right)^{-1}_{\bar{c},n}\left(\tilde{F}^{r}_{\bar{a},a,r}\right)^{-1}_{c,s} (j_{\bar{a},a}^c,j_{c}^{\bar{a},a})
\end{equation}
using the data of both the fusion and module categories. The discussion of cokernels proceeds exactly the same: computing a right inverse is the same calculation with the roles of $a$ and $\bar{a}$ reversed. For the case of non-simple multiplicities the above generalizes straightforwardly: this simply introduces more bulk and boundary junctions to consider (so that the above expressions will contain more indices). 

We further emphasize two points. First, kernels and cokernels may separately trivialize: a left inverse does not imply a right inverse of the same map and vice versa. Second, while these calculations can imply that a subset of kernels and cokernels are trivial, it may be the case the additional kernels or cokernels happen to vanish for non-symmetry related reasons. Such additional vanishings would be model dependent, that is, depend on the specific theory realizing the symmetry. Our calculations show which kernels and cokernels must vanish as a result of the symmetry of the theory. \newline

There are two general cases where the demonstration of \eqref{eq:ker_condition} is particularly straightforward. Suppose $a$ is simple and invertible. For a simple boundary condition $m$, the $\tilde{F}$-symbol then defines an invertible linear map
\begin{equation}
\Hom(\bar{a}\otimes a \otimes m, m) \to \Hom(\bar{a}\otimes a \otimes m, m)
\end{equation}
where
\begin{equation}
    \Hom(\bar{a}\otimes a \otimes m, m) \cong \C.
\end{equation}
Therefore
\begin{equation}
    \left(\tilde{F}^{m}_{\bar{a},a,m}\right)^{-1}_{\mathbf{1} ,a\otimes m} \in \C \setminus \{0\}
\end{equation}
and so the identity channel is non-zero. Moreover since $a$ is invertible, the junction pairing is non-zero only for $c = \mathbf{1}$ and $a\otimes m$ is simple. It follows that for two simple boundaries $m,n$ the map $L_{m,n}^{\bar{a}\otimes m, a\otimes n}(a)$ satisfies \eqref{eq:ker_condition}. Repeating the argument exchanging $a$ and $\bar{a}$ implies both the kernel and cokernel vanish. Therefore for $a$ invertible and $m,n$ simple, $L_{m,n}^{\bar{a}\otimes m, a\otimes n}(a)$ is invertible.

The calculation of kernels and cokernels also simplifies if the boundary $m$ is ``free'' in the module category:
\begin{equation}
    \Hom(m,a\otimes m) \neq 0 \Leftrightarrow a = \mathbf{1}.
\end{equation}
Note that $a$ is no longer required to be invertible. In this case, by definition, all diagrams but the identity channel in \eqref{eq:vanish_ker_condition} will vanish and so it remains only to show that coefficient is non-zero. An argument directly analogous to the above shows this is the case. It follows that if both boundary conditions in the domain are free, the kernel will vanish. Similarly, if both boundary conditions in the codomain are free, the cokernel will vanish.

\subsection{Particle Degeneracies in Massive QFTs}\label{subsec:massive_qft}
We now apply the analysis of Section \ref{subsec:open_sect} to study particle degeneracies in $(1+1)d$ massive QFTs with fusion category symmetry $\cC$. We are interested in stable massive scattering states and so place the theory on $\R^2$. As is familiar when studying scattering on $\R$, particle states are required to resemble clustering ground states at spatial $\pm \infty$. Recall that clustering ground states are those between which local correlators factor 
\begin{equation}
    \bra{\Omega}\mathcal{O}_1(x_1)\mathcal{O}_{2}(x_2)\ket{\Omega} \xrightarrow[|x_1-x_2|\to \infty]{} \braket{\Omega|\mathcal{O}_1(x_1)|\Omega}\braket{\Omega|\mathcal{O}_2(x_2)|\Omega}.
\end{equation}
They are also characterized by the property that they diagonalize local operators in the sense
\begin{equation}
    \braket{\Omega_i|\mathcal{O}(x)|\Omega_j} = \delta_{ij}\braket{\Omega_i|\mathcal{O}(x)|\Omega_i}
\end{equation}
with $\ket{\Omega_i}, \ket{\Omega_j}$ different clustering ground states. This defines the boundary conditions of the theory. The requirement that states resemble clustering ground states at infinity suggests that the behavior of a theory on $\R$ at infinity is governed by its long distance TQFT. We will adopt this perspective. This enables us to deduce properties of the massive theory purely from its infrared behavior. Furthermore, we will think of the theory on $\R$ as the infinite volume limit of a theory on the interval with suitable boundary conditions enforcing clustering in the limit. As a consequence we can use the TQFT on the interval to describe the long distance limit of massive theories.

The theory on $\R^2$ has full Poincar\'{e} symmetry and so we can apply the prior discussion of analyzing degeneracies to single particles states, which furnish irreducible representations of the Poincar\'{e} group. There is a subtlety to consider however. Recall that in $\cH(\R)$ single particle states are very special representations of the Poincar\'{e} group: they are mass eigenstates that are normalizable as wave packets in spatial momenta. It is necessary to consider if the operators constructed from bulk lines and topological boundary junctions preserve this class of representations. This is an infinite volume consideration: the topological line operators have finite action on finite regions of spacetime. Any infinities that arise in their action is governed by their behavior at infinity. This is well controlled however: since these operators have well defined analogues in the infrared TQFT on the interval, they must have finite action on ground states. Therefore the operators in the massive theory must also have finite action at infinity and so they should take normalizable states to normalizable states, and as a result single particle states to single particle states. The degeneracy analysis of the prior section can therefore be used to study degeneracies of stable particles.

In order to do this, we need to know the module category associated to the clustering boundary conditions. The remainder of this section is concerned with this problem. Like fusion categories, module categories enjoy a rigidity property \cite{etingof_tensor_2015} and therefore may in principle be computed directly from the UV if a microscopic model is known. Since the clustering boundary conditions naturally relate to IR data, it is also natural to attempt to extract information from the deep IR instead. We will pursue this approach. A massive QFT with fusion category symmetry $\cC$ will flow to a $\cC$-symmetric TQFT. We briefly review the basic structure of $\cC$-symmetric TQFTs and consider what can be learned about the module category along the flow.

\subsubsection{$\cC$-Symmetric TQFTs}
Given a fusion category $\cC$, a choice of a $\cC$-module category fixes a $\cC$-symmetric TQFT \cite{Huang_2021}. There are related ways to understand this in both $(1+1)d$ and $(2+1)d$. In $(1+1)d$ a choice of $\cC$-module category is the same as the choice of module category of topological boundary conditions of the TQFT.\footnote{More correctly these comments are true on compact manifolds. We comment on the non-compact case below. Note this immediate condition is a strengthening of our previous discussion that boundary conditions produce module categories. In the topological setting, the module category fixes the boundary conditions.} It can be shown constructively that the category of boundary conditions then determines the full $\cC$-symmetric TQFT \cite{Huang_2021}. 

In a $(1+1)d$ TQFT simple boundary conditions (meaning simple objects of the module category) correspond to topological local operators via the open-closed map
\begin{equation}\label{eq:open_closed}
    \includegraphics[width=5.2cm, valign=m]{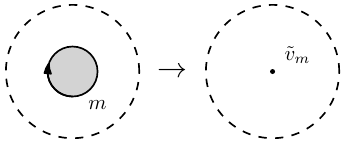}.
\end{equation}
Recall that in TQFTs local operators form an algebra under the OPE. The topological operators given by simple boundary conditions are particularly special with respect to this algebra. Their OPEs can be computed from cutting and gluing \cite{Huang_2021}. More specifically, they follow from the boundary crossing relation
\begin{equation}\label{eq:mod_crossing}
    \includegraphics[width=6.7cm, valign=m]{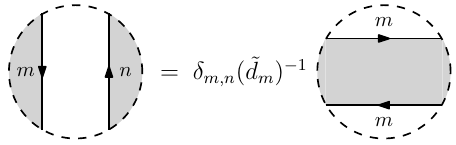}
\end{equation}
where $\tilde{d}_m$ is the ``dimension'' of the boundary condition $m$ in the module category, defined by the diagram
\begin{equation}
    \includegraphics[width=3.1cm, valign=m]{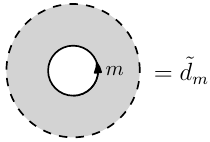}. 
\end{equation}
These diagrams depict a local patch (enclosed in dashed lines) of arbitrary larger diagrams. The crossing \eqref{eq:mod_crossing} follows from gluing the disjoint pieces
\begin{equation}
    \includegraphics[width=11.4cm, valign=m]{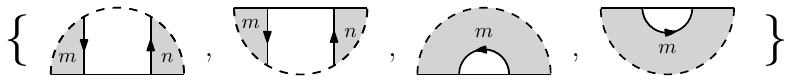}
\end{equation}
in two different ways: one with two and three with four vs one with four and two with three. This gives the algebraic relation
\begin{equation}
    \tilde{v}_m\tilde{v}_n = \includegraphics[width=7cm, valign=m]{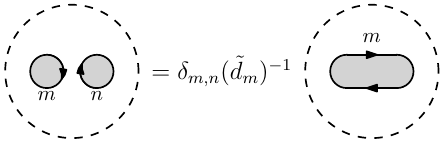} = \frac{\delta_{m,n}}{\tilde{d}_m}\tilde{v}_m.
\end{equation}
Normalizing $v_m = \tilde{v}_m / \tilde{d}_m$ we see
\begin{equation}\label{eq:idempotent}
    v_m v_n = \delta_{mn}v_m.
\end{equation}
Such a basis of topological local operators is called \textit{idempotent}. As in \cite{Bhardwaj:2023idu, Robbins:2022wlr}, we call the topological local operators in this basis ``vacua''. In this language we see that simple boundary conditions correspond to vacua of the TQFT. 

Vacua are also closely related to clustering ground states of the QFT along the RG flow ending in the TQFT. Let $\ket{\Omega_i}$ denote a clustering ground state and $\cH_{\Omega_i,\Omega_j}$ the ground state subspace of the Hilbert space with states resembling $\ket{\Omega_i}$ and $\ket{\Omega_j}$ at $\pm\infty$ respectively. By assumption each sector is gapped and so will end in the TQFT. We additionally make the standard assumption that states in $\mathcal{H}_{\Omega_i,\Omega_i}$ are generated by acting over $\ket{\Omega_i}$. From this it follows that along the flow $\cH_{\Omega_i,\Omega_i}$ has a single ground state corresponding to $\ket{\Omega_i}$. Therefore in the infrared the $\Omega_i$ boundary condition corresponds to a simple topological boundary $m_i$ on the interval. That the boundary is simple follows from the state-operator correspondence on the boundary, defined by the diffeomorphism
\begin{equation}
    \includegraphics[width=5.3cm, valign=m]{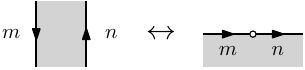}
\end{equation}
and the requirement that simple boundaries support one (up to multiples) topological local operator.

These considerations imply that in the infrared TQFT on $\R$, for each clustering ground state $\ket{\Omega_i}$ there is vacua $v_m$ such that $\braket{\Omega_i|v_m|\Omega_i} \neq 0$. This topological local operator is actually unique as can be seen from the correlator
\begin{equation}\label{eq:clustering_calc}
    \begin{aligned}
        \delta_{m,n} \braket{\Omega_i|v_m|\Omega_i} = \braket{\Omega_i|v_mv_n|\Omega_i} &= \sum_{j}\braket{\Omega_i|v_m|\Omega_j}\braket{\Omega_j|v_n|\Omega_i} \\
        &= \braket{\Omega_i|v_m|\Omega_i}\braket{\Omega_i|v_m|\Omega_i} \\
    \end{aligned}
\end{equation}
where the sum collapses since $v_m(x)$ is a local operator and $\ket{\Omega_i}$ are clustering so that
\begin{equation}
    \braket{\Omega_i|v_m|\Omega_j} = \delta_{i,j}\braket{\Omega_i|v_m|\Omega_i}.
\end{equation}
From \eqref{eq:clustering_calc} two properties immediately follow
\begin{equation}
    \braket{\Omega_i|v_m|\Omega_i} \neq 0 \Rightarrow \braket{\Omega_i|v_m|\Omega_i} = 1
\end{equation}
\begin{equation}
    \braket{\Omega_i|v_m|\Omega_i} = 1 \Rightarrow \braket{\Omega_i|v_n|\Omega_i} = \delta_{n,m}.
\end{equation}
Therefore it follows that to each clustering ground state $\ket{\Omega_i}$ there is \textit{unique} vacua $v_m$ in the infrared TQFT. In particular, this means clustering ground state boundary conditions uniquely correspond to simple topological boundaries of the TQFT on the interval. 

This discussion bolsters the perspective that the behavior of a massive QFT on $\R$ at infinity is governed by its infrared TQFT. We therefore propose using the boundary conditions of the infrared TQFT as the module category of clustering boundary conditions of the QFT along the flow. As remarked above, this is equivalent to knowledge of the infrared TQFT of the QFT itself. It then follows that, given a massive $\cC$-symmetric QFT in $(1+1)d$ on $\R$, knowledge of its infrared $\cC$-symmetric TQFT fixes the possible symmetry enforced particle degeneracies.

\subsubsection{Lagrangian Algebras} \label{LagrangianAlgebraSection}
To briefly summarize. Given knowledge of the module category of boundary conditions of a QFT we have demonstrated both how to deduce the existence of particle degeneracies and that for the case of massive theories on $\R$ this module category is determined by the infrared $\cC$-symmetric TQFT. There remain two practical questions to address. First, how can one systematically characterize the possible infrared behavior of massive $\cC$-symmetric QFTs? Second, and more generally, expressing a module category in terms of topological junctions and crossings for large categories may be difficult and cumbersome to work with. How can they be used efficiently? Both questions may be nicely addressed using the notion of ``algebras''. We now briefly review how this is done, beginning with the second problem. We refer to \cite{kapustin2010surface, Fuchs:2002cm} for further discussion of algebras.

In a fusion category $\cC$ a \textit{Frobenius algebra} $\cA$ is an unoriented, non-simple line $\mathcal{A}$ together with two junctions
\begin{equation}
    \includegraphics[width=1.8 cm, valign=m]{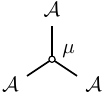} \qquad\qquad \includegraphics[width=1.8 cm, valign=m]{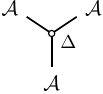}
\end{equation}
which have trivial crossings.\footnote{A Frobenius algebra is also required to come with a ``unit'' and ``counit'' (meaning it contains $\mathbf{1}$ as a line) with their own compatibility conditions. We provide more detail in Appendix \ref{app:lag_alg}.} One can define the notion of a (right) module over $\cA$. Physically this is a bulk line $M$ together with a rule for how $\cA$ can end on it. More concretely it is a pair $(M,p)$ with $M$ a line and $p$ a junction
\begin{equation}
    \includegraphics[width=1.5 cm, valign=m]{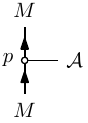}
\end{equation}
having a trivial crossing
\begin{equation}
    \includegraphics[width=4.8 cm, valign=m]{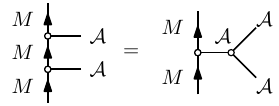}.
\end{equation}
Pairs $(M,p)$ form their own category $\cC_\cA$, with morphisms those in $\cC$ that the line $\cA$ can move past. That is 
\begin{equation}\label{eq:module_morph}
    \includegraphics[width=4.8 cm, valign=m]{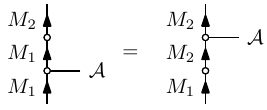}.
\end{equation}
Physically this condition arises naturally viewing $\cC_\cA$ as the collection of interfaces created by gauging algebra $\cA$ in a subset of spacetime, in which case it is a statement of gauge invariance of the topological operators joining the lines.

Algebras and their modules are related to module categories by the following theorem \cite{etingof_tensor_2015}

\begin{theorem}\label{thrm:modcat_alg}
	Take $\cM$ a (left) $\cC$-module category for $\cC$ a fusion category. There exists an algebra $\cA$ in $\cC$ such that $\cM \simeq \cC_{\cA}$ with $\cC_{\cA}$ the category of (right) $\cA$ modules in $\cC$.
\end{theorem}
\noindent This statement is naturally understood in $(1+1)d$, noting that in a TQFT topological boundary conditions can be used to construct topological lines and junctions\footnote{More generally one may construct a bulk topological line from two different topological boundary conditions. Algebraically such a line is called an ``internal hom'' in $\cC$. \cite{etingof_tensor_2015}.}
\begin{equation}
    \includegraphics[width=3.3cm, valign=m]{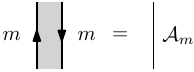}
\end{equation}
\begin{equation}\label{eq:bdry_alg_junction}
    \includegraphics[width=9.1cm, valign=m]{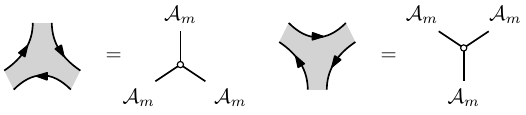}.
\end{equation}
Such junctions can easily be seen to form the Frobenius algebra in Theorem \ref{thrm:modcat_alg}. Algebras therefore provide an efficient way of encoding a module category. In particular the module category $\tilde{F}$-symbols may be obtained by projecting crossings in $\cC$ to topological junctions satisfying \eqref{eq:module_morph}. The characterization of $\cC$-module categories therefore reduces to characterizing algebras in $\cC$. \newline

The module categories of $\cC$ can also be efficiently studied from a $(2+1)d$ perspective. This point of view has the benefit of being particularly constructive. Given a $\cC$-symmetric QFT in $(1+1)d$ there exists a $(2+1)d$ TQFT from which the QFT can be realized by reduction on the interval \cite{Fuchs:2002cm, Fuchs:2003id, Fuchs:2004dz, Fuchs:2004xi, Gaiotto:2014kfa, Gaiotto:2020iye, Apruzzi:2021nmk, Freed:2022qnc, Chatterjee:2022kxb, Inamura:2023ldn, Kaidi:2022cpf, Bhardwaj:2023bbf, Brennan:2024fgj, Antinucci:2024zjp, Bonetti:2024cjk, Apruzzi:2024htg, Apruzzi:2023uma, DelZotto:2024tae}. This TQFT is called the ``symmetry TQFT'' and we will denote it by symTQFT($\cC$). The two boundary conditions used to obtain the $(1+1)d$ theory are called the ``symmetry boundary'' and the ``physical boundary''. The symmetry boundary is topological and supports lines forming $\cC$ while the physical boundary can be non-topological and contains the remaining data of the QFT. Recall that $(2+1)d$ TQFTs are characterized by their topological lines (anyons) which algebraically form an MTC \cite{Kitaev_2006}. The topological lines in symTQFT($\cC$) form $\cZ(\cC)$, the Drinfeld center of $\cC$. Within $\cZ(\cC)$ there is a special class of topological junctions that characterize the topological boundaries of symTQFT($\cC$) called \textit{Lagrangian algebras} \cite{Fuchs:2012dt}.

Physically, a Lagrangian algebra in $\cZ(\cC)$ is a non-simple topological line $L$ together with a topological junction which is both gaugeable and trivializes all bulk anyons when gauged \cite{Kaidi:2021gbs}. By gauging $(L,\mu)$ on a portion of spacetime one obtains a topological boundary condition. Conversely, the Lagrangian algebra is determined by how bulk lines condense on this topological boundary \cite{davydov2011witt,Fuchs:2012dt, Cong:2017ffh, Zhang:2023wlu}. Given a topological boundary, its corresponding Lagrangian algebra can be defined concretely in terms of this condensation. That is, a Lagrangian algebra is a collection of simple lines
\begin{equation}
    L = \oplus_a n_a a, \quad n_a \in \Z_{\geq 0}
\end{equation}
that can end on the topological boundary together with an $M$-symbol defining fusion on the boundary\footnote{We continue writing formulae in the case of simple fusion multiplicities in $\cZ(\cC)$.}
\begin{equation}
    \includegraphics[width=7.5cm, valign=m]{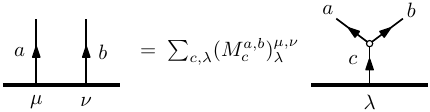}.
\end{equation}
Here the bolded lower boundary represents the two-dimensional topological boundary of symTQFT($\cC$) and the symbols $\{\mu,\nu,\lambda,\dots\}$ denote topological local operators at this junction. The multiplicities $n_a$ give the dimension of the vector space of operators of $a$ ending on the topological boundary. Both $L$ and $M$ must satisfy various constraints that we review in Appendix \ref{app:lag_alg}. 

The Lagrangian algebras of $\cZ(\cC)$ provide a $(2+1)d$ analogue of Theorem \ref{thrm:modcat_alg} \cite{etingof2009weakly, Fuchs:2012dt}:\footnote{In the following indecomposable means the category is not a sum of two smaller categories.}
\begin{theorem}\label{thrm:modcat_lag_alg}
	Take $\cC$ a fusion category. Indecomposable (left) $\cC$-module categories are in bijection with Lagrangian algebras in $\cZ(\cC)$.
\end{theorem}
\noindent This is naturally understood in $(2+1)d$. As previously remarked, a $\cC$-module category is equivalent to a $\cC$-symmetric $(1+1)d$ TQFT. All such TQFTs can be obtained from symTQFT($\cC$) by reduction on the interval with both the symmetry and physical boundary conditions taken to be topological. Suppressing their topological junctions, call the corresponding Lagrangian algebras $L_{sym}$ and $L_{phys}$. Fixing a choice of $L_{sym}$, all other Lagrangian algebras then define the physical Lagrangian of a $(1+1)d$ $\cC$-symmetric TQFT. In this way a choice of Lagrangian algebra in $\cZ(\cC)$ fixes a $\cC$-symmetric TQFT and hence a $\cC$-module category, the statement of the theorem.

In practice Lagrangian algebras provide a particularly convenient, computational tool for studying $\cC$-module categories. Such algebras are highly constrained and as a result the lines that can form one have strong constraints on their multiplicities, dimensions, and spins as we review in Appendix \ref{app:lag_alg}. Moreover, such constraints are typically easy to work with, making finding possible Lagrangian objects straightforward. Given a consistent set of lines, solving for possible $M$-symbols presents the non-trivial aspect of this theory. In practice this is equivalent to solving a pentagon-like polynomial equation \cite{Cong:2017ffh}. However, working with $\cZ(\cC)$, one Lagrangian algebra is always guaranteed to exist: the ``electric Lagrangian''. This Lagrangian is canonically associated to $\cZ(\cC)$ and physically corresponds to the fully spontaneously broken phase of the symmetry. The corresponding topological boundary condition of symTQFT($\cC$) supports lines forming $\cC$ and upon interval reduction produce $\cC$-symmetric TQFT whose boundary module category is the regular $\cC$-module category. This Lagrangian appears in our study of deformations of unitary minimal models in Section \ref{sec:MinModelFlows}. 

For actual calculations a direct relationship between Lagrangian algebras in $\cZ(\cC)$ and $\cC$-module categories is useful. As shown in \cite{etingof2009weakly} this is accomplished simply by condensing the physical Lagrangian algebra $L$ on the symmetry boundary in symTQFT($\cC$).\footnote{In the following we suppress the topological junction defining the algebra.} Doing so $L$ may split as an algebra in $\cC$
\begin{equation}
    L = \oplus_i L_i.
\end{equation}
Choosing any irreducible algebra $L_i$ in the splitting, the corresponding $\cC$-module category is then $\cC_{L_i}$.

\subsubsection{General Procedure} \label{GeneralProcedure}
We summarize this discussion in a general working recipe for studying degeneracies in massive QFTs on $\R$ with non-invertible symmetries in $(1+1)d$. For a theory with symmetry category $\cC$ \newline

\noindent A.  \textit{Determine the module category  $\cM$ of clustering boundary conditions}. This in principle may be determined in different ways. To study it from the infrared $\cC$-symmetric TQFT:
\begin{enumerate}
\item Determine $\cZ(\cC)$.
\item Calculate the Lagrangian algebras in $\cZ(\cC)$ and determine which algebra $L$ characterizes the phase of the QFT.
\item Calculate the projection of this Lagrangian onto the symmetry boundary and fix an indecomposable sub-algebra $L_i \subset L$.
\item From $L_i$ compute the boundary junctions and crossing relations in $\cC_{L_i}$. \newline
 \end{enumerate}

 \noindent B.  \textit{Calculate particle degeneracies}:
 \begin{enumerate}
     \item From the data of $\cM$ and $\cC$ compute \eqref{eq:vanish_ker_condition} for non-zero junctions to determine what kernels or cokernels must vanish. 
     \item Enumerate what boundary sectors are related by injections or surjections. \newline
 \end{enumerate}

This provides a concrete and rather explicit approach to determining the particle degeneracies enforced by simple lines and boundary conditions for a massive QFT having symmetry $\cC$ and boundary module category $\cM$. We now shift to applying this approach study particle degeneracies in massive deformations of unitary minimal models.

%%%%%%%%%%
%%%%%%%%%%

\section{Particle-Soliton Degeneracies in RG Flows from Minimal Models} \label{sec:MinModelFlows}

In this section we study explicit examples of particle degeneracies enforced by non-invertible symmetry. More precisely, we will be interested in RG flows from a CFT to a massive phase triggered by some local relevant operator, in which case the topological line operators preserved along the flow corresponds to those that commute with the bulk local operator triggering the RG flow. It is known that the data of such lines preserved along the flow (such as their fusion ring or $F$-symbols) is rigid \cite{etingof2017fusion, etingof_tensor_2015}, and thus invariant under continuous deformations, in particular RG flows. This observation has been applied mainly to constrain the deep IR of UV CFTs deformed by some symmetry-preserving operator. Here, we use the existence of these topological line operators throughout the RG flow in specific examples in order to illustrate how non-invertible symmetries constrain the particle and soliton spectra of a massive QFT --- in contrast with constraints just in extreme UV or IR limits. To describe these we first start setting some nomenclature on topological line operators in minimal model CFTs. Throughout this section we assume the theories we are considering are unitary. 

\subsubsection*{Verlinde Lines}

We restrict ourselves to diagonal RCFTs with single vacuum, i.e. those with torus partition function
\begin{equation}
    Z_{T^{2}}(\tau, \bar{\tau}) = \sum_{i,j} \delta_{ij} \, \chi_{i}(\tau) \bar{\chi}_{j}(\bar{\tau}),
\end{equation}
where $i,j$ run over the (chiral) primaries of the RCFT, with corresponding characters $\chi_{i}(\tau)$ and modular $S$-matrix $\chi_{i}(-1/\tau) = \sum_{j} S_{ij} \chi_{j}(\tau)$. Throughout this discussion we label the vacuum as $i=0$.

In a diagonal RCFT, a special set of topological line operators with (generically) non-invertible fusion rules are the Verlinde line operators \cite{Verlinde:1988sn,Petkova:2000ip,Drukker:2010jp,Gaiotto:2014lma}. These line operators have the property that they commute with the chiral algebra of the RCFT and are in one-to-one correspondence with the respective (chiral) primaries. Thus, we call $\mathcal{L}_{i}$ and $\phi_{j}$ the Verlinde lines and bulk local operators of the RCFT, respectively, and label them with the same indices. Both satisfy analogous fusion rules:
\begin{equation}
    \phi_{i} \otimes \phi_{j} = \sum_{k} N^{k}_{ij} \, \phi_{k} \, , \qquad \mathcal{L}_{i} \otimes \mathcal{L}_{j} = \sum_{k} N^{k}_{ij} \, \mathcal{L}_{k} \, ,
\end{equation}
where $N^{k}_{ij}$ are the fusion coefficients of the RCFT, from where it is manifest that Verlinde lines satisfy non-invertible fusion rules, generically.

The Verlinde line $\mathcal{L}_{i}$ acts as follows over a bulk local primary operator $\phi_{j}$:
\begin{equation}
    \mathcal{L}_{i}| \phi_{j} \rangle = \frac{S_{ij}}{S_{0j}}| \phi_{j} \rangle.
\end{equation}

Recall that a general topological defect line $\mathcal{L}$ commutes with a bulk local operator $\phi$ if and only if \cite{Chang:2018iay}:
\begin{equation}
    \mathcal{L} | \phi \rangle = \langle \mathcal{L} \rangle | \phi \rangle,
\end{equation}
where $\langle \mathcal{L} \rangle = \langle 0 | \mathcal{L} | 0 \rangle$ is the expectation value of an empty loop of $\mathcal{L}$.\footnote{This expression holds for the expectation value of the line operator on the cylinder. After redefining the line operator by adding a finite local counterterm depending on the extrinsic curvature to absorb the isotopy anomaly we can take it to hold on the plane. See \cite{Chang:2018iay} for more details.} For a Verlinde line, this expectation value is given by the quantum dimension defined in \eqref{eq:dimension}. Specifically, $d_{i} = S_{i0}/S_{00}$ for a Verlinde line $\mathcal{L}_{i}$. Then, the condition for $\mathcal{L}_{i}$ to be preserved along an RG flow triggered by some relevant local operator $\phi_{j}$ is given by
\begin{equation} \label{commutationcondition}
    \frac{S_{ij}}{S_{0j}} = \frac{S_{i0}}{S_{00}}.
\end{equation}
This condition is used below to determine which topological lines are preserved along specific flows from minimal models.

\subsubsection*{Minimal Models}

We now move on to consider the unitary diagonal minimal models (see \cite{DiFrancesco:1997nk} for an overview). Flows from minimal models offer a simple yet rich family of massive QFTs and thus provide a fertile arena to test the formalism outlined in Section \ref{sec:genstructure}. To fix the notation in the following, $M_m$ will denote the minimal model with central charge
\begin{equation}
c = 1 - \frac{6}{m(m+1)} \, , \qquad m \geq 3 \, .
\end{equation}
The spectrum of primary operators $\phi_{r,s}$ can be labelled by the Kac indices $1 \leq r < m , \ 1\leq s < m+1$, where indices related by $(r,s) \leftrightarrow (m-r,m+1-s)$ are identified with the same primary operator. The conformal weights are given by:
\begin{equation} 
h_{r,s} = \frac{\big( (m+1)r - ms \big)^2-1}{4m(m+1)}, \qquad 1 \leq r < m , \qquad 1\leq s < m+1,
\end{equation}
with the same identifications as above. Clearly, $\phi_{1,1}$ corresponds to the vacuum.

The modular S-matrix for $M_{m}$ is given by:
\begin{equation} \label{eq:min_mod_S}
    S_{\rho\sigma,rs} = 2\sqrt{\frac{2}{m(m+1)}}(-1)^{1+s\rho+r\sigma}\sin\left(\pi \frac{m}{m+1}r\rho\right)\sin\left(\pi\frac{m+1}{m}s\sigma\right),
\end{equation}
while the fusion rules are: 
\begin{equation} \label{MinimalModelsFusionRules}
    \mathcal{L}_{rs} \otimes \mathcal{L}_{\rho \sigma} = \sum^{\mathrm{min}(\rho + r - 1, 2m - 1-\rho-r)}_{k = |\rho - r| + 1} \sum^{\mathrm{min}(\sigma + s - 1, 2m+1-\sigma-s)}_{l = |\sigma - s| + 1} \mathcal{L}_{kl} \, ,
\end{equation}
where the summations are constrained such that $(k - \rho + r - 1)$ and $(l - \sigma + s - 1)$ are even.

Using the data just outlined, we can study the Verlinde lines that are preserved when we deform a minimal model by some relevant primary operator $\phi_{r,s}$. Notice that Verlinde lines by definition commute with the chiral algebra, and since the chiral algebra in the case of minimal models is the Virasoro algebra, it follows that all topological defect lines in minimal models are the Verlinde lines.

\subsubsection*{A Small $m$ Example: The Tricritical Ising Model}

For the sake of convenience, we provide a summary of the $M_{4}$ example here. This minimal model is the simplest one that can preserve some non-invertible symmetries along an RG flow and is examined twice below for two distinct flows triggered by two different primaries. 

The spectrum of six bulk local primary operators for $M_{4}$ is summarized in Table \ref{table:tricritical}. Correspondingly, this RCFT has six Verlinde lines. We proceed to describe them and give them particular notations to stress this particular example. One of the Verlinde lines is the identity, which we denote merely by a ``1'' (instead of boldface) as no confusion should arise: $\mathcal{L}_{1,1} \coloneqq 1$. There is a $\mathbb{Z}_{2}$ group-like symmetry in the system given by the line $\mathcal{L}_{3,1}$, which we denote as $\eta \coloneqq \mathcal{L}_{3,1}$ in the following.

The other four Verlinde lines satisfy non-invertible fusion rules. Specifically, we have the line $W \coloneqq \mathcal{L}_{1,3}$, which satisfies Fibonacci fusion rules:
\begin{equation} \label{fibonaccifusionrule}
    W \otimes W = 1 + W \, ,
\end{equation}
and we have the line $N \coloneqq \mathcal{L}_{2,1}$, which along with $\eta$ fulfill a $\mathbb{Z}_{2}$ Tambara-Yamagami fusion ring:
\begin{equation} \label{Z2TYfusionrule}
    \eta \otimes \eta = 1 \, , \qquad \eta \otimes N = N \otimes \eta = N \, , \qquad N \otimes N = 1 + \eta.
\end{equation}
Because of the non-trivial term in the self-fusions of $W$ and $N$, it is clear that such lines are non-invertible. The remaining two lines can be seen as fusion products of the lines already introduced. Explicitly, we have $\mathcal{L}_{2,2} = W \otimes N$ and $\mathcal{L}_{1,2} = W \otimes \eta$. The set of Verlinde lines just described is also summarized in Table \ref{table:tricritical} along with their Kac labels.

\begin{table}[t]
\centering
\begin{tabular}[h]{|p{2.5cm}|p{3.1cm}|p{2.5cm}|p{3.5cm}| }
\hline 
\multicolumn{4}{|c|}{$\mathrm{Tricritical \ Ising \ Model} \ M_{4}$} \\
\hline
Kac label & Conformal Weight & Verlinde Line & Quantum Dimension\\
\hline
$(1,1)$ or $(3,4)$ & $h_{1,1} = 0$ & 1  & $d_{1,1} = 1$ \\
$(1,2)$ or $(3,3)$ & $h_{1,2} = 1/10$ & $W \otimes \eta$ & $d_{1,2} = \frac{1+\sqrt{5}}{2}$ \\
$(1,3)$ or $(3,2)$ & $h_{1,3} = 3/5$ & $W$ &  $d_{1,3} = \frac{1+\sqrt{5}}{2}$ \\
$(1,4)$ or $(3,1)$ & $h_{1,4} = 3/2$ & $\eta$ & $d_{1,4} = 1$ \\
$(2,2)$ or $(2,3)$ & $h_{2,2} = 3/80$ & $W \otimes N$ & $d_{2,2} = \sqrt{2} \big( \frac{1+\sqrt{5}}{2} \big)$ \\
$(2,4)$ or $(2,1)$ & $h_{2,1} = 7/16$ & $N$ & $d_{2,1} = \sqrt{2}$ \\
\hline
\end{tabular}
\caption{Data of the Tricritical Ising Model. On the rightmost column we have summarized the notation we have given to the Verlinde lines associated to the primaries on the leftmost column.}  \label{table:tricritical}
\end{table}

In this example, $M_{4}$ deformed by a single primary $\phi$ preserves some non-invertible symmetry only when $\phi = \phi_{1,3}$ or $\phi =  \phi_{2,1}$. This can be explicitly verified using the modular $S$-matrix expression \eqref{eq:min_mod_S} for $m=4$ and Eqn. \eqref{commutationcondition} above.\footnote{A more general version of this statement for any minimal model flow deformed by $\phi_{1,3}$ can be found in \cite{Kikuchi:2021qxz}. See also Section \ref{phi13arbitraryminimalmodel} below.} To be precise, when we perform the $\phi_{1,3}$ and $\phi_{2,1}$ relevant deformations, the following lines are preserved along the RG flow: 
\begin{align}
    &\phi_{2,1} \ \mathrm{Deformation}: \qquad W \ \mathrm{is} \ \mathrm{preserved}. \label{eq:phi21linespreserved} \\[0.4cm]
    &\phi_{1,3} \ \mathrm{Deformation}: \qquad \eta \ \mathrm{and} \ N \ \mathrm{are} \ \mathrm{preserved}. \label{eq:phi13linespreserved}
\end{align}
All other lines are explicitly broken by these deformations, and any other relevant deformation breaks all the lines with non-invertible fusion rules.

\subsection{Tricritical Ising Deformed by the $\phi_{2,1}$ Operator}\label{fibsec}

The first concrete example we consider is the Tricritical Ising model (i.e. the $M_{4}$ minimal model) deformed by the $\phi_{2,1}$ primary, which we have already described in Section \ref{Introduction}:
\begin{equation}
    S^{2,1}_{M_{4}}[\lambda] = S_{M_{4}} + \lambda \int \phi_{2,1} \, , \quad \lambda > 0,
\end{equation}
The sign of the deformation is actually immaterial, since magnetic operators change by a sign, e.g. $\phi_{2,1} \to - \phi_{2,1}$ under the $\mathbb{Z}_{2}$ symmetry in the UV CFT $M_{4}$, and as such the corresponding deformed Hamiltonians differ by a similarity transformation. The theory flows to a gapped phase in the deep IR \cite{Lassig:1990xy}.

Recall from the discussion in Section \ref{Introduction} that the spectrum of this massive QFT consists of one soliton-antisoliton pair and one particle in the $\cH_{W,W}$ sector \cite{Lassig:1990xy}. Importantly, all these excitations have all the same mass, and in this subsection we set out to recover this degeneracy from the non-invertible symmetry that is preserved along the RG flow.

We move on then to the explicit calculation of the degeneracy. To proceed, we follow the general steps outlined in Section \ref{GeneralProcedure}. Recall from such section that we wish to determine the $\mathcal{C}$-module category describing the clustering boundary conditions at spatial infinity, which may be done by studying the infrared $\mathcal{C}$-symmetric TQFT. In turn, the IR phases can be determined from the Drinfeld center $\mathcal{Z}(\mathcal{C})$ and its Lagrangian algebras (see Appendix \ref{MTCsandalgebras} for definitions). In our case, this will be particularly simple since the fusion category $\mathcal{C}$ preserved along the flow will be a MTC instead of merely a braided fusion category, meaning the Drinfeld center can be written as the double $\mathcal{C} \boxtimes \bar{\mathcal{C}}$. From this expression, it will be straightforward to determine the IR phase and the corresponding module category. Once the module category is determined, we can study the morphisms between Hilbert spaces labeled by different boundary conditions, thus establishing the desired degeneracy between sectors.

Let us apply the procedure just described to the current example. As summarized above, the fusion category $\mathcal{C}$ preserved along this flow is a Fibonacci fusion category consisting of a single non-trivial topological line $W$ with fusion rule \eqref{fibonaccifusionrule}. Additionally, this fusion category can be cast as a MTC (i.e. we can define a non-degenerate braiding),\footnote{This may be checked from the non-degeneracy of the modular $S$-matrix of the Tricritical Ising model restricted to the $(1,1)$ and $(1,3)$ entries on the Kac table (the entries of the lines that are preserved along the flow).} and thus we can study the IR phases by studying the Lagrangian algebras in the Drinfeld double $\mathcal{C} \boxtimes \bar{\mathcal{C}}$, with $\mathcal{C}$ the MTC of Fibonacci anyons consisting of the single non-trivial line $W$ with topological spin $\exp(2 \pi i \frac{3}{5})$. Since the Lagrangian algebra object must consist of simple objects with trivial topological spin, the only allowed Lagrangian algebra in $\mathcal{C} \boxtimes \bar{\mathcal{C}}$ corresponds to the diagonal Lagrangian algebra. It is also easy to check that the conditions \eqref{eq:lag_alg_mult_constraint} and \eqref{eq:lag_alg_dim_constraint} are fulfilled. The IR phase of this massive QFT is thus regular, in the sense that the module category $\mathcal{M}$ we are interested in is described in terms of the data of the fusion category $\mathcal{C}$ itself, as outlined above in Section \ref{subsec:background}. Recall that, physically, the meaning of the module category $\mathcal{M}$ being regular is that the non-invertible symmetry is fully spontaneously broken.

Since the module category we have to consider is regular, the possible boundary conditions are labeled by the simple objects of the Fibonacci fusion category: 1 and $W$, and we can extract the allowed topological junctions between boundaries and bulk lines from the fusion rule \eqref{fibonaccifusionrule}. This means the following topological junctions are allowed:
\begin{equation} \label{FibonacciJunctions}
\vcenter{\hbox{\includegraphics[scale = 0.9]{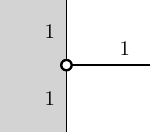}}} \ , \quad \vcenter{\hbox{\includegraphics[scale = 0.9]{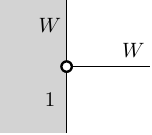}}} \ , \quad 
\vcenter{\hbox{\includegraphics[scale = 0.9]{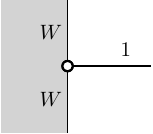}}} \ , \quad \vcenter{\hbox{\includegraphics[scale = 0.9]{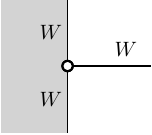}}} \ ,
\end{equation}
and junctions obtained from these ones by reflection through the horizontal are also allowed. Recall that in the regular module category the boundary $\tilde{F}$-symbols coincide with the bulk $F$-symbols, so in the following we use that $\tilde{F} = F$.

From these junctions, one may apply the general arguments outlined in Section \ref{subsec:open_sect} to find mass degeneracies between different sectors. In this example there are so few objects in the categories that all computations may be done explicitly. Indeed, one may straightforwardly arrive at the following morphisms between the different Hilbert spaces:
\begin{equation} \label{GeneralMorphisms}
    \begin{aligned}
        &0 \rightleftarrows \cH_{1,1} \rightleftarrows \cH_{W,W}, \qquad && 0 \rightleftarrows \cH_{1,W} \rightleftarrows \cH_{W,1} \rightleftarrows 0, \\
        &0 \rightleftarrows \cH_{1,W} \rightleftarrows \cH_{W,W}, \qquad && 0 \rightleftarrows \cH_{W,1} \rightleftarrows \cH_{W,W},
    \end{aligned}
\end{equation}
where the different arrows and their directions should be understood as short exact sequences. Thus, for example, the first sequence read from left to right says that there is an injective morphism from $\cH_{1,1}$ to  $\cH_{W,W}$.

To showcase the degeneracies implied by the morphisms \eqref{GeneralMorphisms}, it is actually useful to go through the construction of the morphisms themselves. This is done in the following, with $0 \rightleftarrows \cH_{1,W} \rightleftarrows \cH_{W,W}$ as example.  

From \eqref{FibonacciJunctions}, it is clear we can construct the following morphism:
\begin{equation}
    L_{1,W}^{W,W} \coloneqq \vcenter{\hbox{\includegraphics[scale = 0.9]{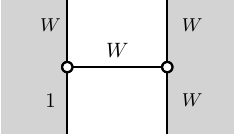}}} \ : \ \cH_{1,W} \to \cH_{W,W} \, ,
\end{equation}
where the compact notation $L_{m,n}^{r,s} \coloneqq L_{m,n}^{r,s}(a)$ is used whenever there is only a single simple bulk line $a$ allowed to perform the morphism from $\mathcal{H}_{m,n} \to \mathcal{H}_{r,s}$ and thus such label is suppressed. A similar notation is used below.

In order to determine whether a state in $\mathcal{H}_{W,W}$ has the same mass as a state in $\mathcal{H}_{1,W}$, we wish to use the morphism $L^{W,W}_{1,W}$ just constructed and apply it to Eqn. \eqref{eq:equivar}. This is not immediate, however, since as discussed in Section \ref{subsec:open_sect} above, \textit{a priori} the morphisms $L_{m,n}^{r,s}$ may have a non-trivial kernel. To show that this is not the case in the current example, consider additionally the following morphism:
\begin{equation}
    L_{W,W}^{1,W} \coloneqq \vcenter{\hbox{\includegraphics[scale = 0.9]{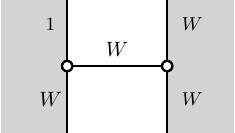}}} \ : \ \cH_{W,W} \to \cH_{1,W} \, .
\end{equation}
We can show that the morphism $L^{W,W}_{1,W}$ ($L^{1,W}_{W,W}$) must be injective (surjective) by demonstrating that it has a left (right) inverse. Consider the composition of the two morphisms, which gives:
\begin{equation}
    L_{W,W}^{1,W} \circ L_{1,W}^{W,W} = \vcenter{\hbox{\includegraphics[scale = 0.9]{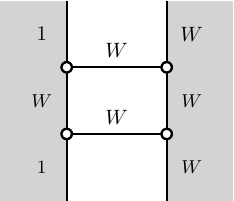}}} \ : \ \cH_{1,W} \to \cH_{1,W} \, .
\end{equation}
This morphism may be simplified by applying the boundary $F$-symbols \eqref{eq:mod_6j}. The only summands with allowed junctions are
\begin{equation}
    L_{W,W}^{1,W} \circ L_{1,W}^{W,W} = c_{1} \ \vcenter{\hbox{\includegraphics[scale = 0.9]{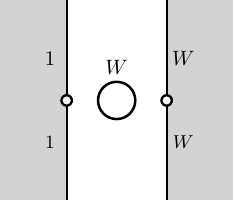}}} + c_{2} \ \vcenter{\hbox{\includegraphics[scale = 0.9]{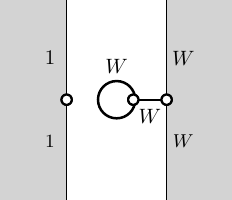}}} \, ,
\end{equation}
where the coefficients $c_1$ and $c_2$ are non-zero and given in terms of the $F$-symbols, but for tidiness we have not written them here explicitly. Also, we have not written the identity line in bulk as it is transparent. 

The key point now is that, recalling the discussion in Section \ref{subsec:open_sect}, the second diagram does not actually contribute. Such vanishing contribution may also be understood in the present context as an instance of the so-called \textit{tadpole vanishing property} of \cite{Chang:2018iay}, which states that any tadpole (sub)diagram of simple lines (other than the identity) enclosing the identity operator must vanish. A quick way to see this general result is to notice that shrinking the loop to a point leads to a putative topological junction between a simple line (here $W$) and the identity line. However, since both lines are simple, no such topological junction exists, and the corresponding diagram has to vanish. All in all, we arrive at the expression
\begin{equation} \label{FibonacciCompositiontoidentity}
    L^{1,W}_{W,W} \circ L^{W,W}_{1,W} = \big[ (F^{1}_{W,W,1})^{-1}_{1,W} \, (F^{W}_{W,W,W})^{-1}_{1,W} \, d_{W} \big] \ \mathrm{Id}_{\mathcal{H}_{1,W}} \, ,
\end{equation}
where we have restored the $F$-symbols and $d_{W}$ is the quantum dimension of the $W$ line. These coefficients are non-zero, so we have successfully shown that $L^{W,W}_{1,W}$ ($L^{1,W}_{W,W}$) has a left (right) inverse and thus that such morphism is injective (surjective).

To see that a multiple of the identity morphism is not always obtained upon using the boundary $F$-symbols, we may instead consider the composition $L^{W,W}_{1,W} \circ L^{1,W}_{W,W}$. Crossing, we pick now a non-zero summand not proportional to the identity:
\begin{equation}
    \vcenter{\hbox{\includegraphics[scale = 0.9]{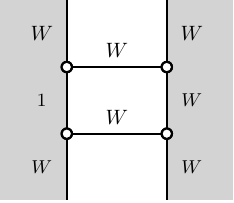}}} = \cdots + (F^{W}_{W,W,W})^{-1}_{W,1} \, (F^{W}_{W,W,W})^{-1}_{W,W} \ \vcenter{\hbox{\includegraphics[scale = 0.9]{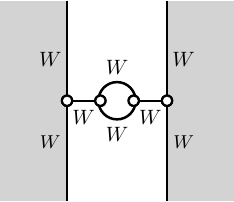}}} \, .
\end{equation}
Thus, we cannot conclude that $L^{1,W}_{W,W}$ has trivial kernel. That is, the non-invertible symmetry does \textit{not} require the $\cH_{W,W} \to \cH_{1,W}$ map to be injective. Indeed, this is nothing but the mathematical realization of the physical fact that $\cH_{W,W}$ has a zero energy state --- the vacuum---  while $\cH_{1,W}$ does not, so the map could not have been injective based on physical grounds from the beginning. The nonexistence of such an injection (equivariant with spacetime symmetry) is thus expected, and provides a non-trivial check of the formalism. This shows that indeed $0 \rightleftarrows \cH_{1,W} \rightleftarrows \cH_{W,W}$.

Now that we have exemplified how the morphisms are obtained, let us return to Eqn. \eqref{GeneralMorphisms}. First, notice $\cH_{1,W}$ always contains a one-particle state.  Indeed, since the boundary conditions are distinct, the minimum energy state must be a stable single particle state.   Starting from the general morphisms \eqref{GeneralMorphisms} then, we see that a (soliton) state of mass $m$ in $\cH_{1,W}$ implies that we necessarily have one (antisoliton) state of the same mass in $\cH_{W,1}$. More interestingly, such a state also implies the existence of one (particle) state of mass $m$ in $\cH_{W,W}$. 

Since $\cH_{1,W}$ and $\cH_{W,1}$ inject into $\cH_{W,W}$ we could have considered the possibility of two states with mass $m$ in $\cH_{W,W}$. To see that this is not the case, consider the composition of injection and surjection $\cH_{W,1} \to \cH_{W,W} \to \cH_{1,W}$. Since $\cH_{W,1}$ and $\cH_{1,W}$ inject into $\cH_{W,W}$, if they formed different vector subspaces of $\cH_{W,W}$ the previous composition would be zero, but this is readily seen to not be the case from the diagrammatic expression
\begin{equation}
    \vcenter{\hbox{\includegraphics[scale = 0.9]{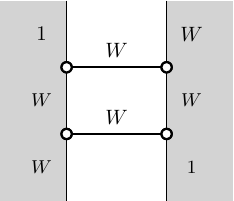}}} = (F^{W}_{W,W,1})^{-1}_{W,W} \, (F^{W}_{W,W,1})^{-1}_{W,W} \ \vcenter{\hbox{\includegraphics[scale = 0.9]{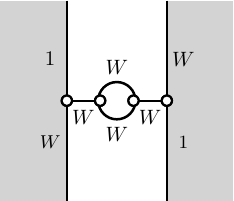}}} \ ,
\end{equation}
where on the right-hand side we have applied the boundary $F$-symbols. Thus, a state with mass $m$ in $\cH_{1,W}$ implies just one state with the same mass in $\cH_{W,W}$ (and another state of the same mass in $\cH_{W,1}$). 

Let us conclude by summarizing our results. Indeed, we have found that when just a Fibonacci fusion category is preserved along an RG flow, a state in $\cH_{1,W}$ implies the existence of a state in $\cH_{W,1}$ and a state in $\cH_{W,W}$ with the same mass. In more standard words, these states form a three-fold multiplet such that they transform in between themselves under the non-invertible symmetry, thus sharing the same mass. Note, that the internal symmetry analysis performed here predicts only that states necessarily come in multiplets of size three. It does not however predict the number of such multiplets apart from the fact that at least one exists.

Specializing this result to the Tricritical Ising model deformed by the $\phi_{2,1}$ primary, we see that this model respects the conclusions above if there is exactly one multiplet, hence reproducing the degenerate spectrum found in \cite{Lassig:1990xy,Zamolodchikov:1990xc} with a corresponding interpretation in terms of solitons and particles. In our case, however, we have recovered this conclusion without having to study the theory in detail, but rather from kinematical considerations based on the existence of a non-invertible symmetry along the RG flow!

\subsection{Tricritical Ising Deformed by the $\phi_{1,3}$ Operator}

We now move on to consider the massive QFT defined by a negative-sign deformation of the Tricritical Ising model by the operator $\phi_{1,3}$:
\begin{equation}
    S^{1,3}_{M_{4}}[\lambda] = S_{M_{4}} - \lambda \int \phi_{1,3} \, , \quad \lambda > 0.
\end{equation}
The negative sign deformation is necessary to find a gapped phase in the IR, which is the physical scenario considered throughout this paper. Otherwise, the RG flow corresponds to the flow between $M_{4}$ and the Ising model $M_{3}$ first studied in \cite{Huse:1984mn,Zamolodchikov:1987ti, Ludwig:1987gs, Zamolodchikov:1991vx}. 

\begin{figure}[!b]
        \centering
        \includegraphics[scale=0.35]{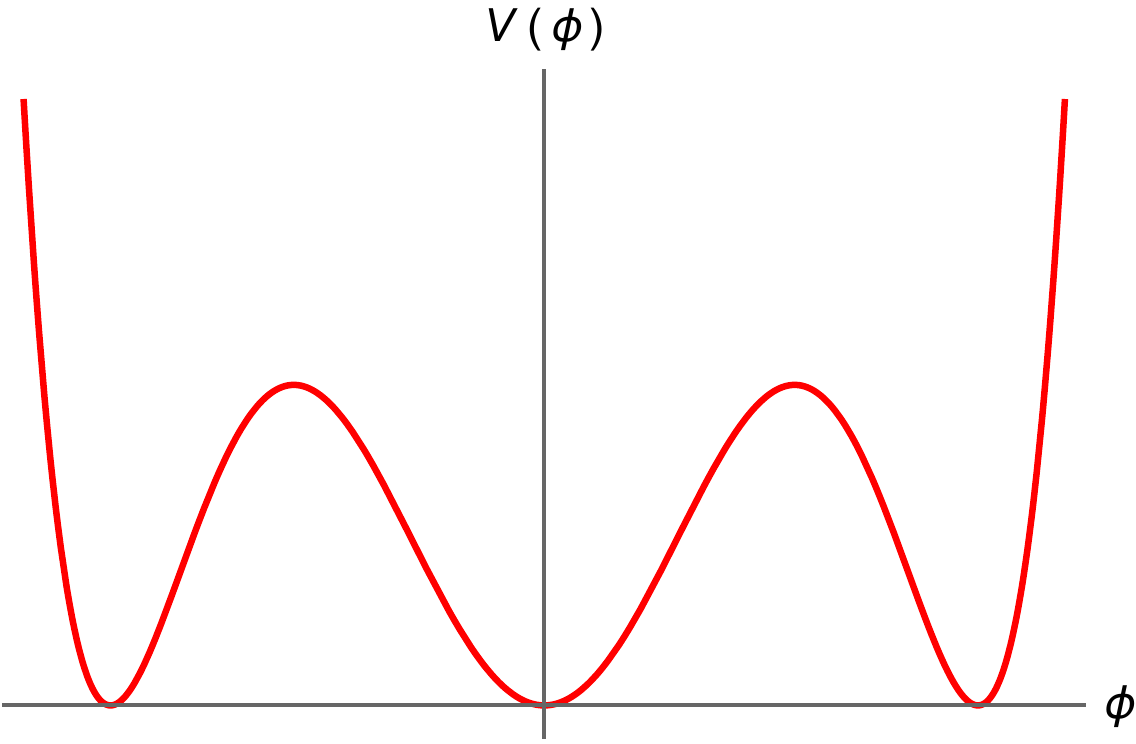} 
        \caption{Schematic depiction of the Landau-Ginzburg potential corresponding to the negative-sign deformation of the Tricritical Ising model deformed by the $\phi_{1,3}$ primary operator. Note in particular the presence of three vacua, in spite of the fact that the only internal group-like symmetry present along the RG flow is $\mathbb{Z}_{2}$.} \label{Tricriticalphi13plot}
\end{figure}

We can motivate the spectrum of this minimal model flow without delving into details ---worked originally in \cite{Zamolodchikov:1991vh}--- by inspecting the associated Landau-Ginzburg realization, whose potential is depicted in Figure \ref{Tricriticalphi13plot} (see Footnote \ref{footnoteLGmodel}). A striking property of this potential is the presence of three vacua (see \cite{Huse:1984mn}), which is interesting since we know from \eqref{eq:phi13linespreserved} that the only group-like topological line present along the flow is a $\mathbb{Z}_{2}$ line, and by itself, it cannot explain on its own this three-fold vacuum degeneracy. The resolution to this puzzle has been beautifully explained in \cite{Chang:2018iay}, where it was argued using modular invariance and --crucially-- the existence of non-invertible topological line operators along the RG flow (see \eqref{eq:phi13linespreserved}) that the number of vacua in the IR TQFT must be (a multiple of) three. Indeed, notice that the duality defect appearing in the $\mathbb{Z}_{2}$ Tambara-Yamagami fusion ring \eqref{Z2TYfusionrule} may be understood as the result of gauging a $\mathbb{Z}_{2}$ symmetry on half of spacetime. In the (gapped) IR, doing this exchanges two vacua on one side of the duality line with a single vacuum on the other side. More precisely, if we call $v_{+}$($v_{-}$) to the rightmost (leftmost) vacuum in Fig. \ref{Tricriticalphi13plot}, and $v_{0}$ the middle one, then the $\mathbb{Z}_{2}$ Tambara-Yamagami lines act over the vacua as follows:

\begin{center}
\begin{tabularx}{0.8\textwidth} 
{ 
  | >{\centering\arraybackslash}X 
  | >{\centering\arraybackslash}X
  | >{\centering\arraybackslash}X
  | >{\centering\arraybackslash}X | }
 \hline
 \ & $v_{0}$ & $v_{+}$ & $v_{-}$ \\
 \hline
1  & $v_{0}$ & $v_{+}$ & $v_{-}$ \\
\hline
$\eta$ & $v_{0}$ & $v_{-}$ & $v_{+}$ \\
\hline
$N$  & $v_{+} + v_{-}$  & $v_{0}$ & $v_{0}$ \\
\hline
\end{tabularx}
\end{center}

\noindent which may be deduced from the consistency of the fusion ring \eqref{Z2TYfusionrule}. As just argued, the duality line $N$ indeed exchanges $v_{\pm}$ with $v_{0}$, in contrast with the $\mathbb{Z}_{2}$ line $\eta$, which just exchanges $v_{+}$ with $v_{-}$ as usual. Thus, we see that the necessary ingredient to ensure the exact three-fold degeneracy of the vacuum is the non-invertible duality line $N$. By the topological nature of $N$, these three vacua must share the same energy. 

Once we understand the threefold degeneracy in terms of the existence of non-invertible topological line defect along the RG flow, it is straightforward to see that four degenerate solitons must exist using the broken $\mathbb{Z}_{2}$ symmetry and CPT. In Figure \ref{Tricriticalphi13plot}, these are the four one-particle solitons extrapolating from either the leftmost or rightmost vacuum to the middle one and vice versa, matching the number of soliton states worked out in \cite{Zamolodchikov:1991vh} from bootstrap considerations. However, since the vacuum structure is determined by the non-invertible symmetries, it is natural to consider if the mass degeneracy of the solitons may be cast directly in terms of the non-invertible symmetries. In the following, we will show that this is indeed the case! Specifically, we will show that the $\mathbb{Z}_{2}$ Tambara-Yamagami symmetry \eqref{eq:phi13linespreserved} preserved along this flow allows for the desired four-fold degeneracy, with the solitons transforming between themselves in the same ``multiplet'' of the non-invertible symmetry. Furthermore, as we will explore in the next subsection, this point of view has the benefit that it extends e.g., to other deformations of minimal models by the $\phi_{1,3}$ operator, where considering all available topological invertible lines and CPT does not explain the full degeneracy. 

We proceed now to the explicit calculation of the degeneracy. The steps to follow have already been described in detail in Section \ref{GeneralProcedure} and summarized for our specific examples in the previous subsection, so we are rather brief in the following. As before, the fusion category preserved along the flow is a MTC, so we can analyze the IR phases by studying the Drinfeld double $\mathcal{C} \boxtimes \bar{\mathcal{C}}$ with $\mathcal{C}$ the $\mathbb{Z}_{2}$ Tambara-Yamagami MTC with fusion rules \eqref{Z2TYfusionrule}. The topological spins may be extracted from Table \ref{table:tricritical} from the lines $(i,1)$, $i=1,2,3$ (those of the Verlinde lines preserved along the flow). \footnote{Notice that the conformal weights of the operators living at the end of lines in the (1+1)d theory generically change along the flow, which can be checked for example for the positive sign deformation by $\phi_{1,3}$ interpolating between the Tricritical Ising model and the Ising model. However, the topological spins of the associated 3D TQFT $\mathcal{C} \boxtimes \bar{\mathcal{C}}$ remain invariant, which may also be explicitly checked in the example just mentioned. Thus, to compute the topological spins in $\mathcal{C} \boxtimes \bar{\mathcal{C}}$ we can choose the conformal weights in $(i,1)$, $i=1,2,3$ lines in Table \ref{table:tricritical} and use Eqn. \eqref{topologicalspinandconformalweight}.} The Lagrangian algebra object must consist of simple objects with trivial topological spin, so the only possible Lagrangian algebra that can be constructed in $\mathcal{C} \boxtimes \bar{\mathcal{C}}$ is the diagonal Lagrangian algebra. The IR phase of this massive QFT is then regular, as in the previous example, with the module category $\mathcal{M}$ described then in terms of the data of the fusion category $\mathcal{C}$ (See Section \ref{subsec:background}).

Since the module category we have to consider is regular, the possible boundary conditions are labeled by the simple objects of the $\mathbb{Z}_{2}$ Tambara-Yamagami fusion category: 1,  $\eta$ and $N$, and we can extract the allowed topological junctions between boundaries and bulk lines from the fusion rule \eqref{Z2TYfusionrule}. This implies that the following topological junctions are allowed:
\begin{align}
& \vcenter{\hbox{\includegraphics[scale = 0.9]{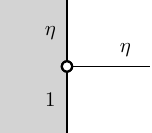}}} \ , \quad \vcenter{\hbox{\includegraphics[scale = 0.9]{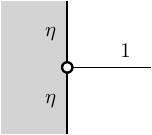}}} \ , \quad 
\vcenter{\hbox{\includegraphics[scale = 0.9]{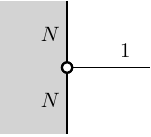}}} \ , \\[0.5cm]
& \vcenter{\hbox{\includegraphics[scale = 0.9]{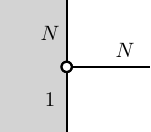}}} \ , \quad \vcenter{\hbox{\includegraphics[scale = 0.9]{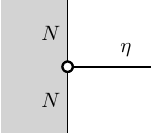}}} \ , \quad 
\vcenter{\hbox{\includegraphics[scale = 0.9]{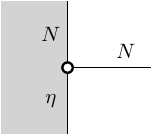}}} \ ,
\end{align}
on top of the one with 1 on all entries, as in the leftmost junction in Eqn. \eqref{FibonacciJunctions}. Junctions obtained from these ones by reflection through the horizontal are also allowed. Recall that in the regular module category the boundary $\tilde{F}$-symbols coincide with the bulk $F$-symbols, so in the following we use that $\tilde{F} = F$.

To show the existence of a four-fold multiplet in between different Hilbert spaces we may refer to the general discussion outlined in Section \ref{subsec:open_sect} to find mass degeneracies between different sectors. But, as in the previous subsection, the categories we consider in specific examples have so few objects that we proceed with the calculations directly. Thus, to show the four-fold multiplet consider first the morphism $\mathcal{H}_{1,N} \to \mathcal{H}_{N,\eta}$ given by diagram
\begin{equation}
    L_{1,N}^{N,\eta} \coloneqq \vcenter{\hbox{\includegraphics[scale = 0.9]{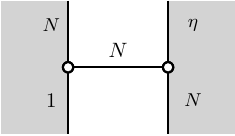}}} \ : \ \cH_{1,N} \to \cH_{N,\eta} \, .
\end{equation}
However, as discussed in Section \ref{subsec:open_sect}, and as in the previous example, we must check that this morphism has a trivial kernel to properly conclude from the equivariance relation \eqref{eq:equivar} that a state of mass $m$ in $\mathcal{H}_{1,N}$ implies the existence of a state of mass $m$ in $\mathcal{H}_{N,\eta}$. To check this, consider the additional morphism
\begin{equation}
    L_{N,\eta}^{1,N} \coloneqq \vcenter{\hbox{\includegraphics[scale = 0.9]{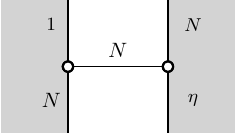}}} \ : \ \cH_{N,\eta} \to \cH_{1, N} \, ,
\end{equation}
and the composition 
\begin{equation}
    L_{N,\eta}^{1,N} \circ L_{1,N}^{N,\eta} = \vcenter{\hbox{\includegraphics[scale = 0.9]{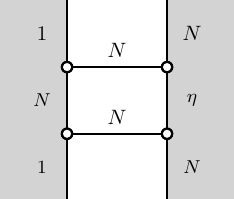}}} \ : \ \cH_{1,N} \to \cH_{1,N} \, ,
\end{equation}
If we show that this composition is proportional to the identity morphism, we would be showing that $L_{1,N}^{N,\eta}$ ($L_{N,\eta}^{1,N}$) in injective (surjective). Thus, we apply the boundary $F$-symbols and write the composition as
\begin{equation}
    L_{N,\eta}^{1,N} \circ L_{1,N}^{N,\eta} = c'_{1} \ \vcenter{\hbox{\includegraphics[scale = 0.9]{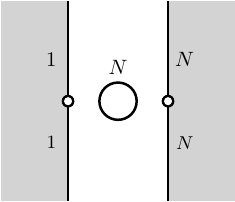}}} + c'_{2} \ \vcenter{\hbox{\includegraphics[scale = 0.9]{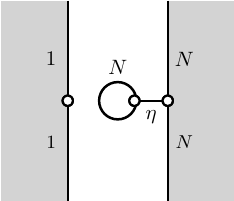}}} \, ,
\end{equation}
where the $c'_{i}$ are quantities given by the $F$-symbols that we do not write explicitly here to avoid clutter. The second diagram in the previous expression vanishes by the vanishing tadpole property (see discussion above \eqref{FibonacciCompositiontoidentity}), so we are left with a single non-vanishing diagram corresponding to the identity morphism. That is, we have:
\begin{equation}
    L_{N,\eta}^{1,N} \circ L^{N,\eta}_{1,N} = [(F^{1}_{N,N,1})^{-1}_{1,N} \, (F^{N}_{N,N,N})^{-1}_{1,\eta} \, d_{N}] \ \mathrm{Id}_{\mathcal{H}_{1,N}} \, .
\end{equation}
Because this composition morphism is proportional to the identity map, we have successfully shown that $L_{1,N}^{N,\eta}$ has a left inverse, and as such $L_{1,N}^{N,\eta}$ is injective. This allows us to conclude via \eqref{eq:equivar} that if there is a state of mass $m$ in $\cH_{1, N}$ there must be a state of mass $m$ in $\cH_{N,\eta}$.

To find the remaining degeneracies, we can consider analogous crossing and vanishing arguments. Specifically, from the following configurations and crossing manipulations:
\begin{equation}
    \vcenter{\hbox{\includegraphics[scale = 0.9]{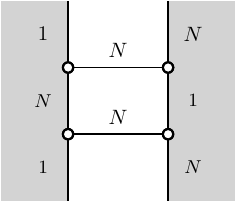}}} = c'_{3} \ \vcenter{\hbox{\includegraphics[scale = 0.9]{images/deformed_cft/1N1NloopwithNloop.pdf}}} + c'_{4} \ \vcenter{\hbox{\includegraphics[scale = 0.9]{images/deformed_cft/1N1Nloopvanishing.pdf}}} \, ,
\end{equation}
and
\begin{equation}
    \vcenter{\hbox{\includegraphics[scale = 0.9]{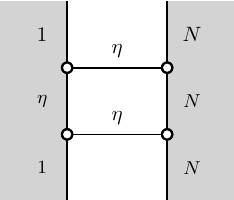}}} = c'_{5} \ \vcenter{\hbox{\includegraphics[scale = 0.9]{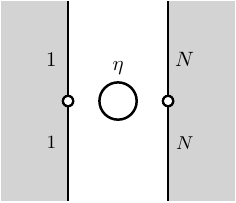}}} \,
\end{equation}
we can conclude that the morphisms $L_{1,N}^{N,1}$ and $L_{1,N}^{\eta,N}$ are injective. It is also straightforward to check that the possible junctions do not allow for additional morphisms from $\cH_{1,N}$ to any of the remaining Hilbert spaces, so the previous manipulations saturate the possible degeneracies given a massive state in $\cH_{1,N}$.

Notice that this last statement is based on the existence of a one-particle soliton state in $\cH_{1,N}$.  The number of such stable one-particle states cannot be a priori determined from symmetry principles alone.  Instead what we conclude is that there must be some number $k\geq 1$ of multiplets of size four in the spectrum.  Analogously, we can consider potential one-particle states in $\cH_{1, \eta}$.  Using this as a seed, a similar argument as in Section \ref{fibsec} demonstrates the existence of a triplet in $\cH_{1, \eta}$, $\cH_{\eta, 1},$ and $\cH_{N,N}$.  Symmetry considerations alone do not tell us the number $\ell \geq 0$ of such triplets.  In summary, any model preserving $\mathbb{Z}_{2}$ Tambara-Yamagami symmetry has integers $k, \ell$ characterizing its single particle spectrum.  Specializing to the Tricritical Ising model deformed by $\phi_{1,3}$ we reproduce the spectrum found in \cite{Zamolodchikov:1991vh} with $k=1$ four particle multiplet and $\ell=0$ three particle multiplets.

\subsection{Minimal Models Deformed by the $\phi_{1,3}$ Operator} \label{phi13arbitraryminimalmodel}

We consider now the generalization of the previous example to a negative-sign deformation of an arbitrary minimal model $M_{m}$ deformed by the least relevant operator $\phi_{1,3}$
\begin{equation}
    S^{1,3}_{M_{m}}[\lambda] = S_{M_{m}} - \lambda \int \phi_{1,3} \, , \quad \lambda > 0.
\end{equation}
As before, we consider the negative sign to have a gapped phase in the IR. The positive sign deformation triggers instead a flow in between successive minimal models as studied in \cite{Huse:1984mn,Zamolodchikov:1987ti, Ludwig:1987gs, Fendley:1993xa}.

The setup in this case is essentially the same as that of the previous example, so we are rather brief. The main difference is that, according to \cite{Zamolodchikov:1991vh}, the number of solitons in the spectrum is now $2(m-2)$ (counting antisolitons). Crucially, all of the solitons are found to have the same mass, and here we reproduce such degeneracy from the non-invertible symmetry present along the flow.

It is straightforward to check that the Verlinde lines that are preserved by the deformation by $\phi_{1,3}$ are those of the form $\mathcal{L}_{i,1}$ with $i=1, \ldots, m-1$ (For the sake of self-containment, we check this in appendix \ref{App:MinimalModelFlows}). Of these lines, only $\mathcal{L}_{m-1,1}$ has non-trivial group-like fusion rules, corresponding to a $\mathbb{Z}_{2}$ symmetry. Notice that in the current case the fusion category of lines preserved along the flow is not necessarily a MTC, so the computation of the Lagrangian algebras is not so trivial. For example, the massive QFT obtained by flowing from the Tetracritical Ising model by $\phi_{1,3}$ has a fusion category that is a braided fusion category, but not modular (see e.g. \cite{Kikuchi:2021qxz}). To proceed, we assume that for an arbitrary minimal model the module category we need is the regular module category. Other than this, the steps are the same as in previous examples. As we will see momentarily, we will be able to reproduce the expected degeneracy from such a setup. In the following, we use a compact notation and denote by ``$n$'' either a boundary condition or a bulk Verlinde line that corresponds to the Kac labels $(n,1)$.

We proceed as in both previous examples. Specifically, we will be interested in showing that 
\begin{equation}
    0 \rightleftarrows \cH_{1,2} \rightleftarrows \cH_{n,n \pm 1} \, ,
\end{equation}
where $n$ is such that $2 \leq n \leq m-2$ or $n = m-1$ as long as we also take the minus sign. We claim that such morphisms will allow us to deduce the expected degeneracy, once we assume a one-particle soliton state in $\cH_{1,2}$. As commented at the end of the previous subsection, other multiplets exist, but we do not study them. Rather, here we concentrate on showing that there exists at least one multiplet implying the $2(m-2)$ expected degeneracy.

Since we are working with an arbitrary minimal model there will be of course multiple allowed junctions, but the ones we will need are:
\begin{equation}
\vcenter{\hbox{\includegraphics[scale = 0.9]{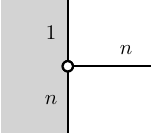}}} \ , \quad \vcenter{\hbox{\includegraphics[scale = 0.9]{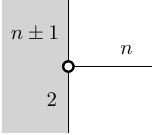}}} \ , \quad 
\vcenter{\hbox{\includegraphics[scale = 0.9]{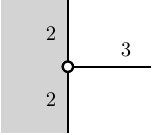}}} \ , \quad \vcenter{\hbox{\includegraphics[scale = 0.9]{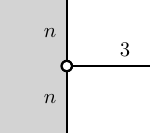}}} \ ,
\end{equation}
on top of the one with 1 on all entries, as in the leftmost junction in Eqn. \eqref{FibonacciJunctions}, and junctions obtained from these ones by reflection through the horizontal are also allowed. The existence of the previous junctions may be readily verified from the fusion rules \eqref{MinimalModelsFusionRules}. From these junctions, we define the following morphisms:
\begin{equation}
    L_{1,2}^{n, n \pm 1} \coloneqq \vcenter{\hbox{\includegraphics[scale = 0.9]{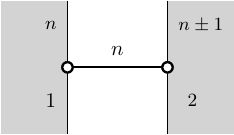}}} \ : \ \cH_{1,2} \to \cH_{n,n \pm 1} \, ,
\end{equation}
and
\begin{equation}
    L_{n, n \pm 1}^{1,2} \coloneqq \vcenter{\hbox{\includegraphics[scale = 0.9]{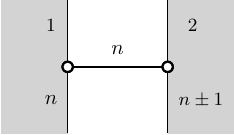}}} \ : \ \cH_{n,n \pm 1} \to \cH_{1,2} \, ,
\end{equation}
 It is straightforward to check that the junctions do not allow for other morphisms from $\cH_{1,2}$ to any other $\cH_{r,s}$. We wish to use $L_{1,2}^{n, n \pm 1}$ in the equivariance relation \eqref{eq:equivar} to show that a state of mass $m$ in $\mathcal{H}_{1,2}$ implies a state of mass $m$ in $\mathcal{H}_{n,n \pm 1}$ for any $n$ in the previous range. As before however, we must check that this map has trivial kernel (see discussion below \eqref{eq:equivar}).

We proceed with the by-now usual argument. We construct the composition map $L^{1,2}_{n, n \pm 1} \circ L^{n,n \pm 1}_{1,2}$ and use the boundary $F$-symbols to find
\begin{equation}
    L_{n, n \pm 1}^{1,2} \circ L_{1,2}^{n, n \pm 1} = c''_{1} \ \vcenter{\hbox{\includegraphics[scale = 0.9]{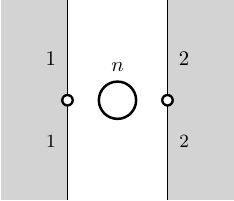}}} + c''_{2} \ \vcenter{\hbox{\includegraphics[scale = 0.9]{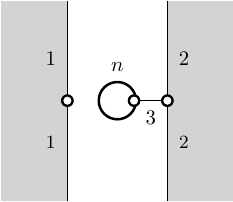}}} \, .
\end{equation}
The second diagram vanishes by the tadpole vanishing property (see discussion above \eqref{FibonacciCompositiontoidentity}), and the loop in the first diagram is the quantum dimension of $n$. Thus, we find
\begin{equation}
    L^{1,2}_{n, n \pm 1} \circ L^{n,n \pm 1}_{1,2} = [(F^{1}_{n,n,1})^{-1}_{1,n} \, (F^{2}_{n,n,2})^{-1}_{1,(n \pm 1)} \, d_{n}] \ \mathrm{Id}_{\mathcal{H}_{1,2}} \, .
\end{equation}
Notice that the second $F$-symbol is non-zero due to the unitarity of the fusion category. Then, the $F$-symbol with those specific entries can be linearly related to the coefficients in \eqref{eq:fusion_dimensions}, which are always non-zero numbers.

Counting the possible morphisms, we see that if we have one state of mass $m_{s}$ in $\cH_{1,2}$, we must have one state of mass $m_{s}$ per allowed morphism above. This gives a multiplet of $(2m-4)$ states with mass $m_{s}$ exchanging states in between different Hilbert spaces, thus reproducing the degeneracy of \cite{Zamolodchikov:1991vh}. As a check, for $m = 4$, we recover a four-fold degeneracy, which is the correct result for the Tricritical Ising model discussed in the previous subsection.

%%%%%%%%%%%%
%%%%%%%%%%%%

\acknowledgments
We are grateful to B. Balthazar for participation at the early stages of this work.  We thank Jimmy Huang, G. Rizi for helpful conversations. CC, DGS and NH acknowledge support from the Simons Collaboration on Global Categorical Symmetries, the US Department of Energy Grant 5-29073, and the Sloan Foundation. DGS is also supported by a Bloomenthal Fellowship in the Enrico Fermi Institute at the University of Chicago.

%%%%%%%%%%%%
%%%%%%%%%%%%

\appendix

\section{Verlinde Lines Preserved by the $\phi_{1,3}$ Operator} \label{App:MinimalModelFlows}

For completeness, we quickly show in this appendix the topological defect lines preserved by RG flows from minimal models triggered by the $\phi_{1,3}$ deformation. Specifically, we are interested in the condition $\cL_{\rho, \sigma} \phi_{1,3}\ket{0} = \phi_{1,3} \cL_{\rho, \sigma} \ket{0}$, or equivalently:
\begin{equation}\label{eq:mod_s_constraint_on_l}
    \frac{S_{\rho\sigma,13}}{S_{11,13}} = \frac{S_{\rho\sigma,11}}{S_{11,11}}.
\end{equation}
From \eqref{eq:min_mod_S}, this constraint is
\begin{equation}
    \frac{\sin\left(\frac{3\pi\sigma}{m}\right)}{\sin\left(\frac{3\pi}{m}\right)} = \frac{\sin\left(\frac{\pi\sigma}{m}\right)}{\sin\left(\frac{\pi}{m}\right)} \Longleftrightarrow \cos{\left(\frac{2 \pi \sigma}{m}\right)} = \cos{\left( \frac{2 \pi}{m} \right)}.
\end{equation}
Thus, the index $\rho$ is unconstrained while $\sigma$ must satisfy this equation in order for $\cL_{\rho\sigma}$ to be preserved. Clearly, the only solution within the allowed range is $\sigma = 1$. So, the $\phi_{1,3}$ perturbed theory retains the $m-1$ topological Verlinde lines given by:
\begin{equation}
    \cL_{\rho, 1} \quad 1 \leq \rho < m.
\end{equation}

\section{MTCs and Algebras} \label{MTCsandalgebras}

In this appendix we summarize some facts about MTCs and Lagrangian algebras, mainly because of its use to pin down the gapped phases we need to consider in Section \ref{sec:MinModelFlows}. Our exposition follows \cite{Cong:2017ffh}, and to simplify the discussion we restrict the formulae to the case of simple multiplicities at the junctions.

\subsection{Braidings, Twists, MTCs}
We are interested in the topological defect lines (anyons) in $(2+1)d$ topological quantum field theories, which simple lines we denote as $a, b, c, \ldots$ throughout this appendix. In this case the fusion category of lines comes equipped with the additional data of a braiding, encoded in the $R$-symbol
\begin{equation}\label{eq:rmatrix}
    \includegraphics[width=4.5cm, valign=m]{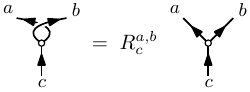}.
\end{equation}
The $R$-symbol is required to satisfy a set of consistency conditions whose precise form will not be necessary for our purposes \cite{Kitaev_2006}. 

From the braiding, one can define the topological spin $\theta_{a}$ of a line $a$ as:
\begin{equation}
    \theta_a \coloneqq \sum_{c}\frac{d_c}{d_a} R^{a,a}_c.
\end{equation}
Geometrically, the topological spin captures the effect of twisting a line by $2\pi$.\footnote{That this process is non-trivial reflects that the topological lines are really framed.} When the TQFT lives as the bulk supporting a $(1+1)d$ rational CFT on its boundary, the topological spins of the topological lines are related to the boundary data as
\begin{equation} \label{topologicalspinandconformalweight}
    \theta_{a} = e^{2 \pi i h_{a}},
\end{equation}
where $h_{a}$ corresponds to the conformal weight of the primary associated to the Verlinde line obtained by pushing $a$ to the boundary.

Given a braiding one can further define the linking of simple lines
\begin{equation}\label{eq:rmatrix}
    \includegraphics[width=3.5cm, valign=m]{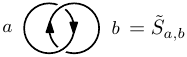}.
\end{equation}
If the matrix $\tilde{S}_{a,b}$ is non-degenerate, the fusion category is then called a \textit{modular tensor category} (MTC). Roughly, this reflects the fact that any line can be detected by a non-trivial braiding with some other line.

\subsection{Lagrangian Algebras}\label{app:lag_alg}

In unitary theories, a computationally concrete characterization of Lagrangian algebras can be given. Physically, this definition encodes how bulk lines of symTQFT($\cC$) (see Section \ref{LagrangianAlgebraSection}) end on the topological boundary. 

A \textit{Lagrangian algebra} is a non-simple line $L = \oplus_a n_a a$ such that
\begin{equation}\label{eq:lag_alg_mult_constraint}
    n_a n_b \leq \sum_c N_{a,b}^{c} \, n_c \, , \qquad n_1 = 1,
\end{equation}
and
\begin{equation}\label{eq:lag_alg_dim_constraint}
    \left(\sum_{a \in L} n_a d_a\right)^2 = \sum_{b\in \cZ(\cC)} d_b^2 \, ,
\end{equation}
together with an $M$-symbol defined by the diagram
\begin{equation}
    \includegraphics[width=7.5cm, valign=m]{images/general_degen/general_TFT_Msymbol.pdf}.
\end{equation}
The bolded lower boundary represents the topological boundary of the $(2+1)d$ TQFT, and the indices $\{\mu,\nu,\lambda,\dots\}$ denote distinct possible topological junctions for a given bulk line ending on the boundary. Such junctions form vector spaces (boundary condensation spaces) at the end of each simple line in $L$. 

The $M$-symbol is required to satisfy two polynomial equations:
\begin{equation}\label{eq:msymbol_crossing}
\sum_{\sigma} \left(M^{a,b}_{e}\right)^{\mu,\nu}_{\sigma} \left(M_d^{e,c}\right)^{\sigma,\lambda}_\phi  = \sum_{f,\psi}
\left(M^{b,c}_{f}\right)^{\nu,\lambda}_{\psi} \left(M_d^{a,f}\right)^{\mu,\psi}_\phi F_{\bar{c},\bar{b},\bar{a}}^{\bar{d};\bar{f},\bar{e}} \, ,
\end{equation}
\begin{equation}\label{eq:msymbol_braiding}
    \left(M^{b,a}_c\right)^{\nu,\mu}_{\lambda} R^{a,b}_c = \left(M^{a,b}_c\right)_{\lambda}^{\mu,\nu},
\end{equation}
where we have used the $F$ and $R$ symbols for the braided fusion category $\cZ(\cC)$ here. These algebraic conditions encode the following geometric properties of the bulk topological lines in symTQFT($\cC$)
\begin{equation}
     \includegraphics[width=7cm, valign=m]{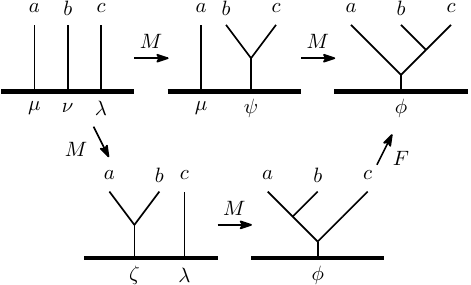}
\end{equation}
and
\begin{equation}
     \includegraphics[width=4.5cm, valign=m]{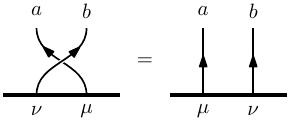}.
\end{equation}
Moreover, \eqref{eq:lag_alg_mult_constraint} expresses that $M$ is not required to be unitary (unlike $F$ and $R$).

As equations \eqref{eq:lag_alg_mult_constraint} and \eqref{eq:lag_alg_dim_constraint} show, the simple lines that form a Lagrangian algebra must fulfill some constraints. Indeed, an additional constraint on top of \eqref{eq:lag_alg_mult_constraint} and \eqref{eq:lag_alg_dim_constraint}, is that the simple lines forming the Lagrangian algebra must be bosons. That is:
\begin{equation} \label{VanishingTopologicalspin}
    n_a \neq 0 \Longrightarrow \theta_a = 1.
\end{equation}
A physical way of understanding this constraint is by thinking on the topological boundary as arising from gauging (in a generalized sense) the Lagrangian algebra on half of spacetime (see e.g., the discussion in \cite{Kaidi:2021gbs}). For instance, when the gauging corresponds to condensing abelian anyons, the above condition corresponds to the vanishing of the 't Hooft anomaly of the one-form symmetry, encoded in turn in the topological spin of the topological lines of the $(2+1)d$ theory \cite{Gaiotto:2014kfa, Gomis:2017ixy, Hsin:2018vcg}. As such, the anyons can be gauged on half of spacetime and a topological boundary is developed. The constraint \eqref{VanishingTopologicalspin} may be seen as a more general version of this requirement, and also holds when the components of the Lagrangian algebra are not necessarily abelian anyons.

In practice, conditions \eqref{eq:lag_alg_mult_constraint}, \eqref{eq:lag_alg_dim_constraint} and \eqref{VanishingTopologicalspin} are highly constraining and easy to work with. These conditions do not imply, however, the existence of a compatible $M$-symbol solving \eqref{eq:msymbol_crossing} and \eqref{eq:msymbol_braiding}, which in practice are very difficult equations to solve. Moreover, the same Lagrangian algebra object $L$ may admit distinct $M$-symbols. The construction of a consistent $M$ therefore encompasses the non-trivial aspect of rigorously studying Lagrangian algebras. In this work, we mostly consider the constraints \eqref{eq:lag_alg_mult_constraint}, \eqref{eq:lag_alg_dim_constraint} and \eqref{VanishingTopologicalspin}, which are sufficient to study the gapped phases arising in the flows considered in Section \ref{sec:MinModelFlows}. Recall from the discussion in Section \ref{LagrangianAlgebraSection} that a Lagrangian algebra in $\mathcal{Z}(\mathcal{C})$ is always guaranteed to exist, the so-called ``electric Lagrangian.'' In particular, when $\mathcal{Z}(\mathcal{C}) = \mathcal{C} \boxtimes \bar{\mathcal{C}}$ with $\mathcal{C}$ some MTC, such a Lagrangian algebra is the diagonal Lagrangian algebra of the form
\begin{equation}
    L = \oplus_{a} (a,\bar{a}),
\end{equation}
where $a$ runs over the simple anyons of $\mathcal{C}$ and $\bar{a}$ denotes its counterpart in $\bar{\mathcal{C}}$, where $\bar{\mathcal{C}}$ is the orientation reversal of $\mathcal{C}$.

\bibliographystyle{JHEP}
\bibliography{references}

\end{document}